\documentclass[11pt]{article}

\usepackage{acl}
\usepackage{booktabs}

\usepackage{times}
\usepackage{latexsym}
\usepackage{amsmath}
\usepackage{amssymb}
\usepackage{array}
\usepackage{tabularx}
\usepackage{enumitem}

\usepackage[T1]{fontenc}

\usepackage[utf8]{inputenc}

\usepackage{microtype}

\usepackage{inconsolata}

\usepackage{graphicx}
\usepackage{subcaption}

\usepackage{arydshln}

\usepackage{multirow} 
\usepackage{xcolor}

\usepackage[labelformat=simple]{subcaption}

\title{Teaching LLMs to Learn Tool Trialing and Execution through Environment Interaction}

\author{
    Xingjie Gao$^{1}$\thanks{ \ \ indicates equal contribution.}, 
    Pengcheng Huang$^{1}$\footnotemark[1], 
    Zhenghao Liu$^{1}$\thanks{ \ \ indicates corresponding author.} \\ 
    \textbf{Yukun Yan$^{2}$, Shuo Wang$^{2}$, Zulong Chen$^{3}$, Chen Qian$^{4}$, Ge Yu$^{1}$, Yu Gu$^{1}$} \\
    \\
    $^1$School of Computer Science and Engineering, Northeastern University, Shenyang, China \\
    $^2$Department of Computer Science and Technology, Tsinghua University, Beijing, China \\
    $^3$Alibaba Group, Hangzhou, China \\
    $^4$School of Artificial Intelligence, Shanghai Jiao Tong University, Shanghai, China \\
}

\newcommand{\method}{ToolMaster}

\begin{document}
\maketitle

\begin{abstract}
Equipping Large Language Models (LLMs) with external tools enables them to solve complex real-world problems. However, the robustness of existing methods remains a critical challenge when confronting novel or evolving tools. Existing trajectory-centric paradigms primarily rely on memorizing static solution paths during training, which limits the ability of LLMs to generalize tool usage to newly introduced or previously unseen tools. In this paper, we propose \method{}, a framework that shifts tool use from imitating golden tool-calling trajectories to actively learning tool usage through interaction with the environment. To optimize LLMs for tool planning and invocation, \method{} adopts a trial-and-execution paradigm, which trains LLMs to first imitate teacher-generated trajectories containing explicit tool trials and self-correction, followed by reinforcement learning to coordinate the trial and execution phases jointly. This process enables agents to autonomously explore correct tool usage by actively interacting with environments and forming experiential knowledge that benefits tool execution. Experimental results demonstrate that \method{} significantly outperforms existing baselines in terms of generalization and robustness across unseen or unfamiliar tools. All code and data are available at \url{https://github.com/NEUIR/ToolMaster}.
\end{abstract}


\section{Introduction}
\begin{figure}[t] 
\centering
    \includegraphics[width=\linewidth]{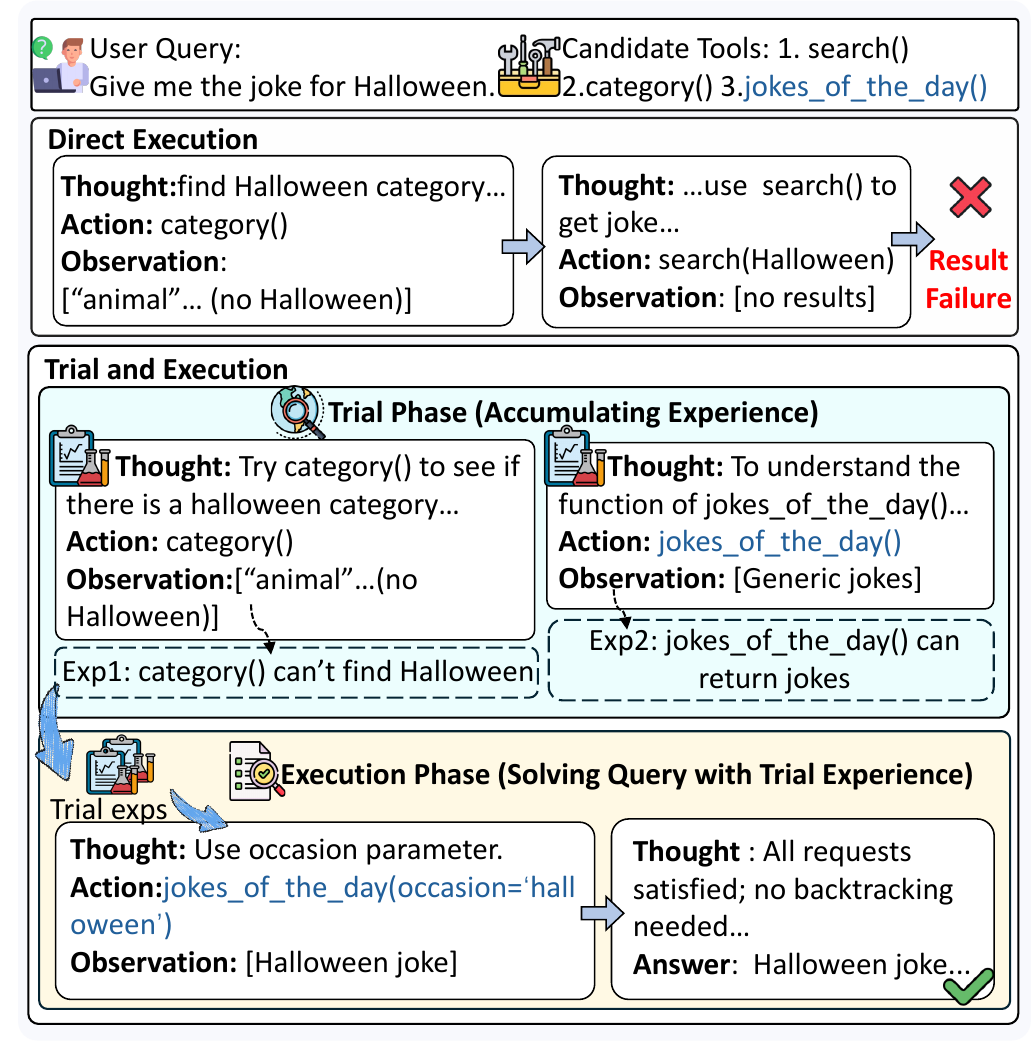}
    \caption{Illustration of the Trial-and-Execution paradigm proposed by \method{}.} \label{fig:pre-expla}
\end{figure} 

Large Language Models (LLMs) have demonstrated strong planning and reasoning capabilities, and equipping them with external tools has proven effective in further enhancing their ability to solve complex, real-world problems~\cite{qin2023toolllm,restgpt,qin2024}. Typical examples include using calculators to improve numerical accuracy~\cite{cobbe2021training, parisi2022talm} and leveraging search engines to retrieve factual knowledge~\citep{carlini2021extracting, thoppilan2022lamda, borgeaud2022improving}. To further enhance the tool-use capability of LLMs, recent studies have primarily focused on enabling LLMs to perform more effective tool planning, tool decision, and reliable tool invocation~\citep{shen2023hugginggpt, lu2023chameleon, liang2023taskmatrix}.

Earlier works~\cite{yao2023react,restgpt} mainly rely on prompting-based methods to decompose tool-learning tasks into sub-tasks and generate grounded plans by leveraging the reasoning capabilities of LLMs~\cite{wei2023cot}. However, such approaches remain significantly limited in their tool-use performance~\cite{qin2023toolllm}. To overcome these limitations, recent research on tool-using agents has increasingly adopted a trajectory-centric post-training paradigm~\cite{qin2023toolllm,tang2023toolalpaca,qian2025toolrl,feng2025retool,steptool}. The dominant approach typically involves collecting high-quality trajectories through sampling methods (e.g., MCTS) or expert demonstrations, followed by Supervised Fine-Tuning (SFT)~\cite{qin2023toolllm} or Reinforcement Learning (RL)~\cite{qian2025toolrl} to enforce imitation of desired trajectories.
While effective on fixed toolsets, this trajectory-centric paradigm brute-forces LLMs to imitate specific tool-use trajectories, causing them to struggle when tools evolve or when deployment scenarios deviate from these supervisions~\cite{zeng-etal-2025-tool-zero}.

To enhance the tool-use generalization capability of LLMs, existing methods typically leverage their self-reflection and self-asking abilities to enable more accurate tool planning and invocation~\cite{mekala2024toolverifier, ToolMVR}.
However, a fundamental challenge remains: LLMs often lack robustness in real-world applications when explicit feedback from environments is unavailable~\cite{wang2024intervenor}. As illustrated in Figure~\ref{fig:pre-expla}, when an LLM is presented with a newly introduced tool ``jokes\_of\_the\_day()'', which is required to solve the task, alongside a well-learned tool ``search()'' that fails to return relevant knowledge, the model may still directly invoke ``search()'', leading to incorrect results. This behavior likely arises because ``search()'' appears frequently in the training data, making the LLM overly confident in invoking it. In contrast, by benefiting from tool trials that enable tool-usage experiments through interactions with the environment, the LLM is able to perform correct tool invocation and obtain accurate results, highlighting the necessity of tool trialing in tool learning tasks.

In this paper, we build \method{} upon the Trial-and-Execution paradigm, aiming to fully exploit tool-calling feedback from the environment prior to tool planning and invocation. Specifically, \method{} first optimizes LLMs via supervised fine-tuning to imitate tool-trialing behaviors using tool-calling trajectories generated by a teacher model. Subsequently, we employ reinforcement learning to further optimize the model, enabling it to jointly coordinate tool trialing and tool execution actions for more accurate outcomes.
During tool-calling trajectory synthesis, we adopt a more capable LLM as the teacher model and prompt it to perform tool trialing by invoking tools to obtain feedback, followed by the tool execution phase that leverages the accumulated experiences for explicit tool planning and self-correction.

Our experiments on three different tool-learning datasets demonstrate the effectiveness of \method{}, which achieves more than a 7\% improvement over baseline models. Benefiting from the trial phase, LLMs perform more tool-calling interaction steps, which significantly reduces execution failures and leads to higher accuracy. Moreover, \method{} exhibits strong generalization capability, reflected in its accuracy in both unfamiliar tool-calling scenarios and problem-solving that requires previously unseen tools. Notably, \method{} also alleviates unnecessary biases when invoking tools that are rarely observed in the training dataset.

\section{Related Work}

Tool use extends the capabilities of Large Language Models (LLMs) by allowing them to interact with external environments. Early paradigms for tool use, such as RestGPT~\cite{restgpt} and ReAct~\cite{yao2023react}, relied on in-context learning to prompt LLMs to leverage tools for problem solving. To further enhance tool use capabilities, ToolLLM~\cite{qin2023toolllm} employs supervised fine-tuning of LLMs using the collected dataset ToolBench, which contains large-scale tool usage trajectories constructed via the Depth-First Search-based Decision Trees (DFSDT) method. However, such SFT-based methods that rely on curated trajectories tend to overfit the training signals and suffer from catastrophic forgetting~\citep{luo2023empirical}.

Instead of SFT, recent works have further employed reinforcement learning methods to optimize LLMs to enhance their capabilities in tackling complex tool-use tasks. TP-LLaMA~\cite{tp-llama} applies Direct Preference Optimization (DPO)~\cite{rafailov2024direct} to align models with preferred tool paths, as well as ToolRL~\cite{qian2025toolrl} and Tool-Zero~\cite{zeng-etal-2025-tool-zero} further leverage the Group Relative Policy Optimization (GRPO) method~\cite{shao2024grpo} to optimize LLMs, emphasizing the design of sophisticated reward functions for guidance, such as the accuracy of tool calls along ground-truth trajectories. Furthermore, to ensure the effectiveness of training, FTRL~\cite{ye2025ftrl} proposes a stable and verifiable method for synthesizing tool-use training data.
However, these methods may suffer from the reward hacking problem~\cite{skalse2022defining}, where the model performs fewer trials to maximize the tool-calling accuracy reward, thereby limiting their generalization across different tool-use scenarios~\cite{mekala2024toolverifier}.

To enhance the generalization ability of LLMs in tool use, substantial efforts have been directed towards enhancing their capability to use new tools and incorporating the tool execution feedback for self-correction. Some works~\cite{mekala2024toolverifier} employ self-asking contrastive questions for tool selection and parameter generation to fully exploit new tools. Other methods~\cite{ToolMVR} introduce self-correction mechanisms by reflecting on errors in the tool-calling trajectories based on feedback from the tool executions. 
However, these approaches mainly focus on correcting tool-calling errors using environmental feedback, neglecting the proactive agentic role of LLMs in autonomously conducting tool-calling trials for planning and reasoning.


\section{Methodology}
\begin{figure*}[t] 
\centering
    \includegraphics[width=1.0\textwidth]{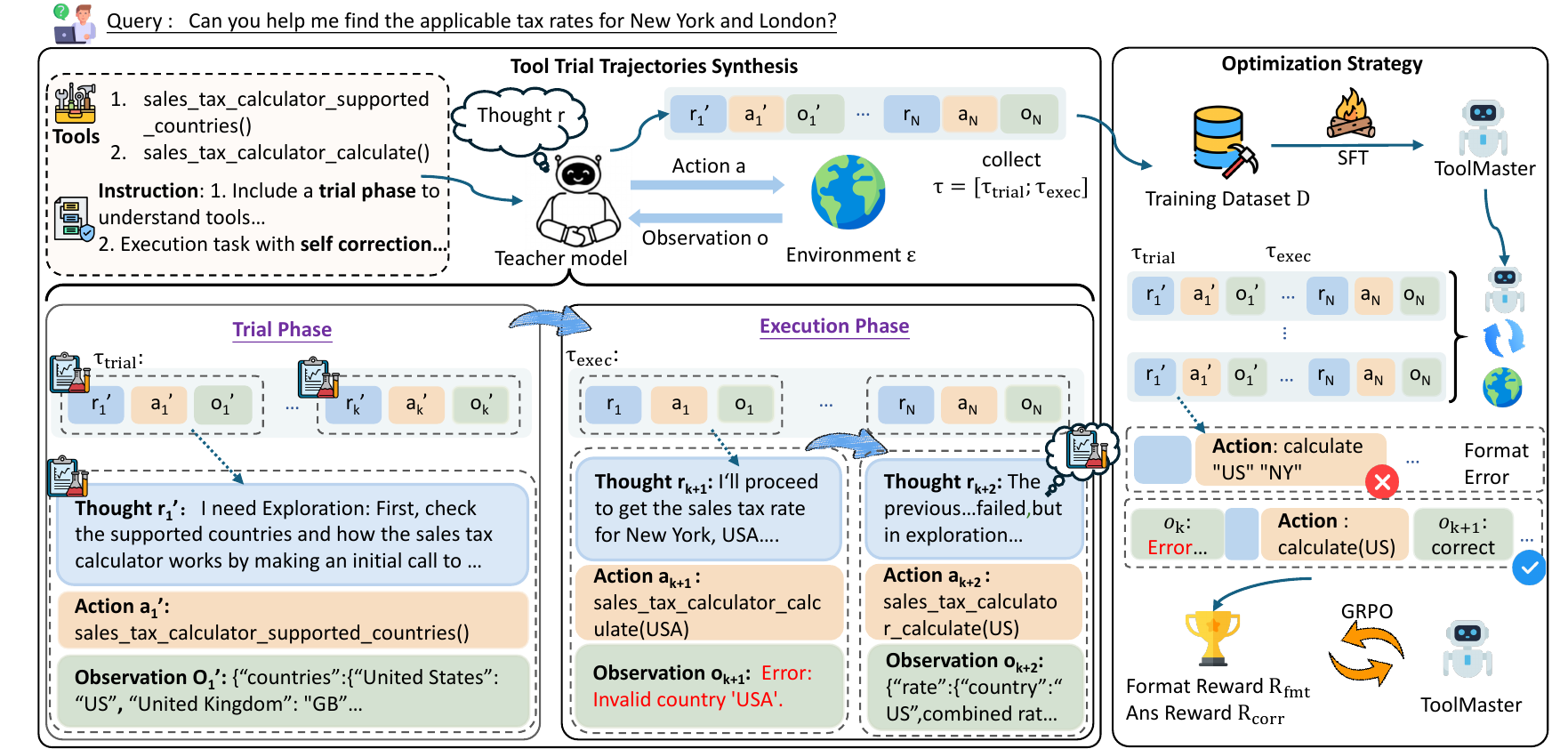}
    \caption{The architecture of \method{}. } \label{fig:method}
\end{figure*}


In this section, we present \method{}, a framework illustrated in Figure~\ref{fig:method} that optimizes LLMs to master tools through trial-and-execution paradigm.
Specifically, the model first conducts pre-interactions in the trial phase to accumulate tool-usage experience and refine its internal belief. After this initial calibration, the model proceeds to the execution phase, where it leverages the tool-usage experience to iteratively solve the task through self-correction. To realize this paradigm, we first introduce the optimization pipeline (Sec.~\ref{sec:optimize}), detailing the training strategy used to instill such reasoning capabilities. To facilitate this training, we then detail the underlying data synthesis methodology (Sec.~\ref{sec:data_construction}), explaining how we prompt LLMs to generate the requisite high-quality tool trial trajectories.

\subsection{Optimizing LLM Tool Exploration Capability via Environment Interaction}
\label{sec:optimize}
To enhance the tool planning capability, we optimize LLMs to perform more effective tool exploration by conducting trials to execute tools within the environment $\mathcal{E}$.

\textbf{Tool Planning with Environment Feedback.} Given a user query $q$ and a set of candidate tools $\mathcal{S}_{\text{Tool}}$, the tool learning task requires the LLM to perform tool planning, invoke appropriate APIs, and interact with the external environment to obtain an intermediate reasoning result to generate the final answer $y$:
\begin{equation}
\small
\tau, y = \text{LLM} (q, \mathcal{S}_{\text{Tool}}),
\end{equation}
where the reasoning trajectories $\tau$ can be represented:
\begin{equation}
\small
\tau = \{(r_1, a_1, o_1), (r_2, a_2, o_2), \dots, (r_N, a_N, o_N)\},
\end{equation}
where $r_i$ denotes the reasoning step at time $i$, $a_i \in \mathcal{S}_{\text{Tool}}$ represents the action to trigger an API call with structured arguments, and $o_i$ is the observation returned by the environment after executing $a_i$.

At each step $i$, the LLM conditions on the accumulated context to produce reasoning and actions:
\begin{equation}
\small
p(r_i, a_i \mid q, h_{i-1}),
\end{equation}
where the history $h_{i-1} = \{(r_j, a_j, o_j)\}_{j=1}^{i-1}$ includes all previous reasoning steps, tool calls, and observations. 
The environment executes the selected tool action and returns an observation:
\begin{equation}
\small
o_i = \mathcal{E}(a_i),
\end{equation}
which is appended to the context for subsequent reasoning.
This iterative reasoning-action-observation loop enables the model to decompose complex queries into a sequence of grounded tool invocations, allowing explicit interaction with the environment $\mathcal{E}$. However, such a tool planning paradigm does not fully exploit the feedback signals returned by the environment during tool invocation, as it fails to explicitly incorporate essential environment feedback into the tool planning process.

\textbf{Trajectory Optimization Strategies.} To enhance the tool exploration capability of LLMs, we first optimize the model to acquire richer tool exploration behaviors by distilling reasoning trajectories from a superior LLM. Subsequently, Group Relative Policy Optimization (GRPO)~\cite{shao2024grpo} is employed to maximize execution success within the environment $\mathcal{E}$.

First, \method{} prompts a superior LLM to generate tool trial trajectories by explicitly instructing tool trials and self-correction behaviors, thereby constructing the SFT dataset $\mathcal{D}$ (Sec.~\ref{sec:data_construction}).
For each trajectory $\tau \in \mathcal{D}$, we optimize the LLM parameters $\theta$ by minimizing the following loss:
\begin{equation}
\small
\mathcal{L}_{\text{SFT}} = -\mathbb{E}_{\tau \sim \mathcal{D}} \left[ \sum_{i=1}^{N} \log \pi_\theta(r_i,a_i \mid q,h_{i-1}) \right].
\end{equation}
We then adopt a composite outcome-based reward function to maximize the expected cumulative reward during LLM optimization:
\begin{equation}
\small
\mathcal{L}_{\text{GRPO}} = -\mathbb{E}_{\tau \sim \pi_\theta} [R_{\text{fmt}} + R_{\text{corr}}],
\end{equation}
where $R_{\text{fmt}} \in \{0, 1\}$ denotes the format reward, which is set to 1 if the model strictly follows the prescribed reasoning and tool invocation schema, and $R_{\text{corr}} \in \{0, 1\}$ denotes the answer correctness reward, which is assigned a value of 1 if the final answer correctly resolves the user query. To estimate the answer correctness, we employ a stronger LLM as an automatic judge to evaluate the tool execution results and compute the reward $R_{\text{corr}}$.

\subsection{Prompting LLMs to Synthesize Tool Trial Trajectories for SFT}
\label{sec:data_construction}
In this subsection, we describe the methodology for constructing the SFT dataset $\mathcal{D}$ by synthesizing high-quality training trajectories that autonomously encourage LLMs to explore tool usage under the \textit{trial-and-execution} paradigm.
Specifically, we employ a strong reasoning-capable model, such as DeepSeek-V3.1~\cite{deepseekv3}, as the teacher $\pi_{\text{teacher}}$, and prompt it to fully engage in tool-usage trials while collecting rich feedback from tool invocations. The feedback from the environment is incorporated as a part of the tool-use experience, which is leveraged to enhance both tool planning and task-solving capabilities.

\textbf{Tool Trialing with Environment.}
In the trial phase, the model fully interacts with the environment to collect sufficient feedback on tool usage:
\begin{equation}
\small
\tau_{\text{trial}}  \sim \pi_{\text{teacher}}(\cdot \mid \mathcal{I}, q, \mathcal{S}_{\text{Tool}}),
\end{equation}
where $\mathcal{I}$ is the instruction. The trajectory $\tau_{\text{trial}}$ contains $k$ autonomous tool-calling trials, which are determined by $\pi_{\text{teacher}}$:
\begin{equation}
\small
   \tau_{\text{trial}} = \{(r_j^\prime, a_j^\prime, o_j^\prime)\}_{j=1}^{k},
\end{equation}
where each tuple $(r_j^\prime, a_j^\prime, o_j^\prime)$ corresponds to an investigative step rather than a direct solution attempt. Specifically, the reasoning thought $r_j^\prime$ formulates a hypothesis to explore particular tool semantics or parameter constraints; the tool action $a_j^\prime$ executes a probing operation to verify functional behaviors; and the observation $o_j^\prime$ reveals the environmental feedback during tool invocation. Through these interactions, the model accumulates grounded observations with the trajectory $\tau_{\text{trial}}$ as empirical experience, allowing it to calibrate its understanding of the available toolset within the actual environment prior to the tool planning and execution phase.

\textbf{Tool Execution.} Given the tool-use experience $\tau_{\text{trial}}$, the model generates an execution trajectory $\tau_{\text{exec}}$ for tool planning and invocation:
\begin{equation}
\small
\tau_{\text{exec}} \sim \pi_{\text{teacher}}(\cdot \mid  \mathcal{I}, q, \mathcal{S}_{\text{Tool}}, \tau_{\text{trial}}),
\end{equation}
where the problem-solving trajectory $\tau_{\text{exec}}$ contains $N$ tool invocations to resolve the query $q$:
\begin{equation}
\small
\tau_{\text{exec}} = {(r_j, a_j, o_j)}_{j=1}^N,
\end{equation}
where each $(r_j, a_j, o_j)$ denotes an execution step. Specifically, $r_j$ analyzes the current context to formulate a solution strategy, $a_j$ executes a purposeful operation to advance the task, and $o_j$ represents the feedback from the environment.
To enable the teacher model to effectively leverage error signals from the environment, we incorporate a self-correction mechanism that rectifies the intermediate state, ensuring that the trajectory is guided back toward the correct final answer.
Consequently, the final answer $y$ is given by:
\begin{equation}
\small
 y \sim \pi_{\text{teacher}}(\cdot \mid  \mathcal{I}, q, \mathcal{S}_{\text{Tool}}, \tau_{\text{trial}}, \tau_{\text{exec}}).
\end{equation}

\textbf{SFT Data Curation.} Finally, we collect the SFT dataset $\mathcal{D}$ for SFT. Formally, for each query $q$, we construct a trial-and-execution trajectory $\tau$ by concatenating the sub-trajectories from the trial and execution phases:
\begin{equation}
\small
\tau = [\tau_{\text{trial}}; \tau_{\text{exec}}].
\end{equation}
Subsequently, the final SFT dataset $\mathcal{D}$ is obtained by filtering for high-quality trajectories:
\begin{equation}
\small
\mathcal{D} = \{ (q_1, \tau_1, y_1), \dots, (q_K, \tau_K,y_K) \},
\end{equation}
where only the trajectories $\tau$ that successfully resolve the corresponding query $q$ and strictly adhere to the behavioral guidelines specified in $\mathcal{I}$ are retained. Further details of the filtering methodology are provided in Appendix~\ref{sec:Details of Implementation}.

\section{Experimental Methodology}

This section describes datasets, baselines, and implementation details used in experiments.

\textbf{Datasets.}
To construct our training data, we leverage the training split of the publicly available ToolBench~\cite{qin2023toolllm}. Specifically, we curate a subset of 1,500 queries to construct the SFT dataset and 800 queries for RL training.
To ensure a comprehensive evaluation, we employ one in-domain benchmark: (1) StableToolbench~\cite{guo2025stabletoolbench}, a stabilized suite covering diverse domains and multi-tool compositions; and two Out-of-Domain (OOD) benchmarks to assess generalization: (2) TMDB~\cite{song2023restgpt}, which tests precise API mapping and argument filling, and (3) ToolHop~\cite{ye2025toolhop}, which evaluates complex multi-hop reasoning and cross-tool planning.


\textbf{Baselines.} 
For a comprehensive evaluation, we benchmark \method{} against three distinct categories of baselines: (1) Zero-shot LLMs, (2) SFT-based baselines, and (3) RL-based methods.

First, for Zero-shot LLMs, we evaluate powerful proprietary and open-weights models, specifically GPT-4o~\cite{gpt-4o}, GPT-4o-mini~\cite{gpt4o-mini}, and Qwen2.5-32B-Instruct~\cite{qwen2.5}. Additionally, we report the zero-shot performance of the backbone models employed in our training to serve as a direct baseline for quantifying improvement.
Regarding SFT-based Baselines, we compare against Distill (SFT), which is fine-tuned on successful tool-use trajectories distilled from DeepSeek-V3.1~\cite{deepseekv3}, and ToolLLM~\cite{qin2023toolllm}, a robust data-centric approach that utilizes a Depth-First Search Decision Tree (DFSDT) to construct high-quality solution paths for instruction tuning.
Finally, we compare against several RL-based methods, including StepTool~\cite{steptool}, FTRL~\cite{ye2025ftrl}, and ToolRL~\cite{qian2025toolrl}. Specifically, FTRL~\cite{ye2025ftrl} and ToolRL~\cite{qian2025toolrl} are built upon the GRPO framework with different reward formulations, whereas StepTool~\cite{steptool} optimizes the policy with PPO and assigns step-wise rewards to enable explicit reward assignment for each intermediate tool-use step.


\textbf{Evaluation Metrics.} Following previous work~\citet{codetool}, for StableToolBench~\cite{guo2025stabletoolbench} and TMDB~\cite{song2023restgpt}, we adopt the Solvable Pass Rate (SoPR) for evaluation. Following \citet{ToolMVR}, we leverage GPT-4o as the evaluator and utilize the same prompts to categorize responses into ``Solved'', ``Unsolved'', or ``Unsure''. A score of 1 is assigned to ``Solved'' instances, while others receive 0.
For ToolHop~\cite{ye2025toolhop}, we evaluate Answer Correctness based on whether the model's output contains the ground truth answer. More details of evaluation prompts and criteria are provided in Appendix~\ref{sec:More Details on Experiment}.

\begin{table*}[!t]
    \centering
    \small

    \begin{tabular}{llccccccc} 
    \toprule
    \textbf{Backbone Model} & \textbf{Method} & \textbf{I1 Inst} & \textbf{I1 Cat} & \textbf{I1 Tool} & \textbf{I2 Cat} & \textbf{I2 Inst} & \textbf{I3 Inst} & \textbf{Avg.} \\ 
    \midrule
    
    \multirow{1}{*}{GPT-4o} & Zero-shot & 59.28 & 56.21 & 64.56 & 61.29 & 60.38 & 60.66 & 60.23 \\
    \multirow{1}{*}{GPT-4o-mini} & Zero-shot & 56.44 & 52.29 & 65.19 & 54.84 & 50.00 & 50.82 & 54.93 \\
    \multirow{1}{*}{Qwen2.5-32B-Instruct} & Zero-shot & 54.60 & 52.29 & 56.96 & 55.65 & 48.11 & 50.82 & 53.07 \\

    \midrule
    
    \multirow{2}{*}{LLaMA-2-7b-hf} 
    & ToolLLM & 50.92 & 43.14 & 51.92 & 41.94 & 39.62 & 42.62 & 45.03 \\
    & StepTool & 39.26 & 38.56 & 41.67 & 30.58 & 34.91 & 31.15 & 36.02 \\

    \midrule

    \multirow{6}{*}{Qwen2.5-7B-Instruct} 
    & Zero-shot  & 49.08 & 44.44 & 50.63 & 41.94 & 36.79 & 37.70 & 43.43 \\
    & Distill (SFT) & {55.83} & {56.21} & 56.96 & {54.84} & 50.00 & 49.18 & 53.84 \\
    & ToolLLM & 52.15 & 54.25 & 55.70 & 50.81 & 47.17 & \underline{57.38} & 52.91 \\
    & ToolRL & 52.15 & 49.67 & 51.90 & 44.35 & 49.06 & 44.26 & 48.57 \\
    & FTRL & \underline{60.43} & \underline{57.19} & \underline{62.66} & \underline{59.27} & \underline{54.25} & 56.56 & \underline{58.39} \\
    & \method{} & \textbf{66.26} & \textbf{65.36} & \textbf{65.82} & \textbf{66.13} & \textbf{64.15} & \textbf{70.49} & \textbf{66.37} \\
    
    \midrule

    \multirow{3}{*}{Qwen3-8B} 
    & Zero-shot & 50.92 & 53.59 & 53.80 & 45.97 & 40.57 & 40.98 & 47.64 \\
    & Distill (SFT) & 60.12 & \underline{60.13} & \underline{65.19} & 52.42 & \underline{64.15} & 52.46 & 59.08 \\
    & FTRL & \textbf{64.42} & 58.82 & 62.03 & \underline{61.29} & 53.77 & \underline{55.74} & \underline{59.34}\\
    & \method{} & \underline{63.19} & \textbf{64.71} & \textbf{66.46} & \textbf{66.94} & \textbf{67.92} & \textbf{68.85} & \textbf{66.34} \\

    \midrule

    \multirow{3}{*}{Qwen3-14B} 
    & Zero-shot & 55.83 & 55.56 & 61.39 & 51.61 & 46.23 & 57.38 & 54.67 \\
    & Distill (SFT) & \underline{66.26} & 63.40 & \underline{68.99} & 59.68 & 58.49 & 49.18 & 61.00 \\
    & FTRL & 60.12 & \underline{64.05} & 62.66 & \underline{62.10} & \underline{59.43} & \underline{59.02} & \underline{61.23} \\
    & \method{} & \textbf{69.33} & \textbf{67.32} & \textbf{72.78} & \textbf{70.97} & \textbf{75.47} & \textbf{70.49} & \textbf{71.06} \\ 
    
    \bottomrule
    \end{tabular}
    
\caption{Overall performance comparison of different methods on StableToolBench. The best results are highlighted in \textbf{bold}, and the second-best results are \underline{underlined}.}
    \label{tab:overall} 
\end{table*}

\textbf{Implementation Details.} 
We conduct experiments on three backbone models: Qwen2.5-7B-Instruct~\cite{qwen2.5}, Qwen3-8B~\cite{qwen3}, and Qwen3-14B~\cite{qwen3}. For data synthesis, we employ DeepSeek-V3.1~\cite{deepseekv3} as the teacher to generate SFT trajectories, and subsequently utilize it as a verifier to perform data filtering for quality assurance. Regarding the training configuration, we first train the models in the SFT phase for 3 epochs with a learning rate of $1 \times 10^{-5}$ and a maximum sequence length of 8,192. In the subsequent GRPO stage, we adopt a learning rate of $1 \times 10^{-6}$, set the KL coefficient to $\beta=0.002$, and utilize a group size of 4 with the correctness reward $R_{\text{corr}}$ determined by DeepSeek-V3~\cite{deepseekv3} based on its direct evaluation of the execution trajectorys task fulfillment. During inference, we set the temperature to 0.1 for all models to ensure consistent evaluation. Detailed implementations are reported in Appendix~\ref{sec:Details of Implementation}.

\section{Evaluation Results}
In this section, we first present the overall performance of \method{}. Subsequently, we demonstrate its generalization capabilities in out-of-domain (OOD) settings, followed by ablation studies and analyses to validate the effectiveness of our proposed trial-and-execution paradigm.

\begin{table}[!t]
    \centering
    \small
    \begin{tabular}{lccc} 
    \toprule 
    \textbf{Method} & \textbf{TMDB} & \textbf{ToolHop} & \textbf{Avg.}   \\ 
    \midrule 

    GPT-4o-mini  & 75.00 & 42.21 & 57.60 \\
    GPT-4o & 80.00 & 45.32 & 61.81  \\
    Qwen2.5-32B-Instruct & 73.00 & 20.00 & 46.50   \\
    \midrule 
    \multicolumn{3}{l}{\textit{Qwen2.5-7B-Instruct}} \\
    \hdashline
    Zero-shot & \underline{69.00} & 16.68 & 42.84   \\
    Distill (SFT) & 65.00 & 27.43 & 46.22   \\
    ToolLLM (\citeyear{qin2023toolllm}) & 68.00 & \underline{33.96} & \underline{50.98}  \\
    FTRL (\citeyear{ye2025ftrl}) & 56.00 & 29.05 & 42.53 \\
    ToolRL (\citeyear{qian2025toolrl}) & 67.00 & 32.46 & 49.73 \\
    
    \method{} & \textbf{86.00} & \textbf{37.38} & \textbf{61.69}  \\ 
    \midrule 
    
    \multicolumn{3}{l}{\textit{Qwen3-8B}} \\
    \hdashline
    Zero-shot & 79.00 & 38.59 & 58.80  \\ 
    Distill (SFT) & 70.00 & \underline{43.71} & 56.86 \\
    FTRL (\citeyear{ye2025ftrl}) & \underline{81.00} & 40.30 & \underline{60.65} \\
    \method{} & \textbf{82.00} & \textbf{45.03} & \textbf{63.52}  \\ 
    
    \bottomrule 
    \end{tabular}
\caption{Performance on TMDB and ToolHop to evaluate the generalization capability of different methods.}
    \label{tab:generalization-results} 
\end{table}
\subsection{Overall Performance}
This subsection shows the overall performance of \method{} under both in-domain and out-of-domain testing settings.

We first present the main results of \method{} across all subsets of StableToolBench in Table~\ref{tab:overall}.
Overall, \method{} consistently outperforms all baseline methods, achieving an average improvement of more than 7\% over baseline models. Notably, this advantage remains stable across different backbone models, demonstrating the generalization ability of \method{}. Compared with prompting-based methods, \method{} yields over 16\% improvements, indicating that relying solely on prompting LLMs to enable tool use capability is less effective. In comparison with Distill (SFT), \method{} also achieves substantial gains, demonstrating the benefit of adopting more effective training strategies, such as RL methods, to better leverage supervision signals for guiding LLMs in tool usage. Furthermore, when compared with RL-based methods such as FTRL and ToolRL, \method{} shows improvements at least 7\%, underscoring the effectiveness of the trial-and-execution paradigm, which constructs valuable tool-use experiences through iterative interaction with tools and then utilizes the experiences for tool planning and invocation.

To further validate the generalization capability of \method{}, we conduct evaluations under out-of-domain (OOD) settings, specifically assessing performance in tool-rich environments (TMDB) and complex multi-hop reasoning scenarios (ToolHop). As shown in Table~\ref{tab:generalization-results}, \method{} consistently outperforms all baselines, surpassing the strongest competitor by an average margin of 6.8\%. This notable performance gain demonstrates the strong generalization ability of \method{} in unseen tool use scenarios. We attribute this advantage to the trial-based experiments, which enable the model to accumulate tool-use experience by actively testing tools in the environment. Such experience equips the model with more transferable tool-use knowledge, allowing it to effectively solve problems in previously unseen environments.

\subsection{Ablation Study}
\label{sec:ablation}
In this subsection, we conduct ablation studies to evaluate the contribution of various components in \method{} and report the results across different difficulty levels (I1, I2, I3) of StableToolBench.

As shown in Table~\ref{tab:ablation-study}, we first examine the impact of different training strategies used by \method{}. Specifically, \method{} w/o SFT and \method{} w/o RL remove the SFT and RL processes, respectively, to assess their effectiveness. \method{} w/o Trial-and-Exec uses only the golden tool-use trajectories, omitting the trial-and-execution process during SFT training. Next, we investigate the role of the trial-and-execution mechanisms in constructing the SFT dataset. \method{} (SFT) w/o Trial Phase removes the tool trial process, while \method{} (SFT) w/o Self-Correction eliminates the self-correction step within the execution process. Additionally, \method{} (SFT) w/o Trajectory Filter is included to demonstrate the impact of the data filtering strategy during SFT dataset construction.

\begin{table}
    \centering
    \small 
    
    \resizebox{\linewidth}{!}{%
    \begin{tabular}{lcccc}
    \toprule 
    \textbf{Method} & \textbf{I1} & \textbf{I2} & \textbf{I3} & \textbf{Avg.} \\ 
    \midrule 
    
    \method{} & \textbf{65.81} & \textbf{65.15} & \textbf{70.49} & \textbf{66.37} \\ 
    w/o SFT & 60.41 & 57.06 & 55.74 & 58.52 \\
    w/o RL\ & 59.70 & 60.90 & 57.38 & 59.72 \\
     w/o Trial-and-Exec & 62.00 & 61.16 & 60.66 & 61.50 \\

    \midrule 
    \method{} (SFT)\ & 59.70 & 60.90 & 57.38 & 59.72 \\
    w/o Trial Phase & 56.73 & 54.78 & 52.46 & 55.37 \\
    w/o Self-Correction & 57.58 & 56.67 & 50.82 & 56.15 \\
    w/o Trajectory Filter & 56.05 & 57.13& 55.73& 56.36 \\
    \bottomrule 
    \end{tabular}}

    \caption{Performance of components in \method{}. All models are implemented using Qwen2.5-7B-Instruct.}
    \label{tab:ablation-study} 
\end{table}

The evaluation results show that, when removing the SFT or RL training phase, the performance of \method{} degrades, demonstrating the necessity of both. Notably, the application of the Trial-and-Execution paradigm yields an additional improvement of approximately 5\% (comparing \method{} with \method{} w/o Trial-and-Exec), as it enables the model to conduct tool-use trials, forming experiments that facilitate tool planning and invocation. Next, we analyze the roles of different components within the trial and execution phases in curating the SFT dataset. The results indicate that both the trial phase and self-correction within the execution phase effectively supervise LLMs for tool usage. The trial phase is particularly beneficial for easier tasks (I1 and I2), as it helps form effective experiments that improve tool-use accuracy, while self-correction proves more effective for more challenging tasks (I3), stimulating self-reflection to verify and refine the reasoning process. Furthermore, removing the data filter leads to performance degradation, validating the necessity of training on high-quality, filtered trajectories.

    

\begin{figure}[t]
    \centering
    \begin{subfigure}{0.48\linewidth}
        \centering
        \includegraphics[width=\linewidth]{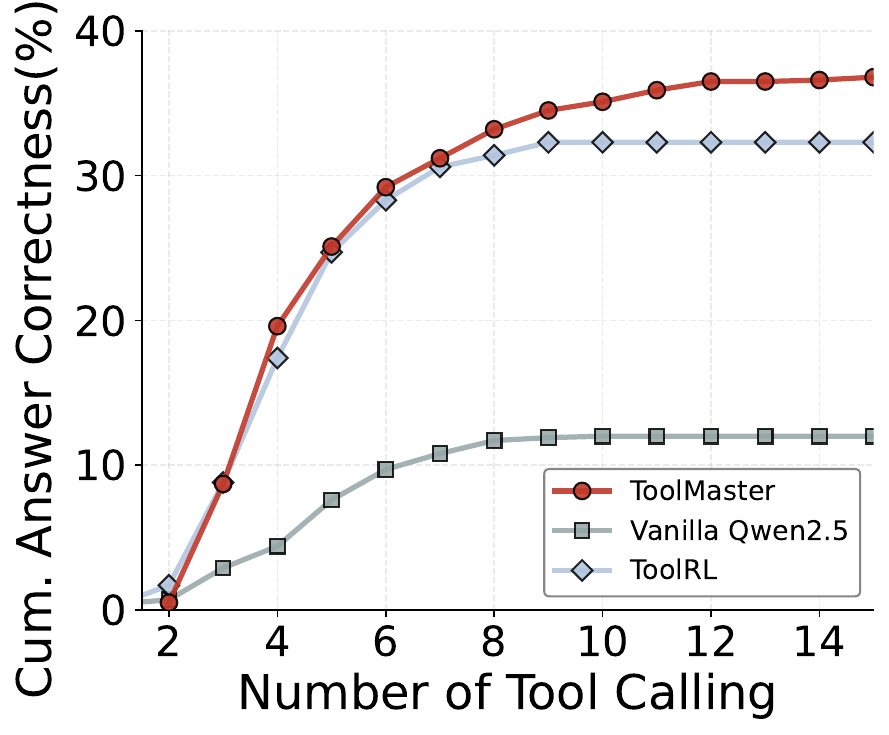}
        \caption{The correlation between answer correctness and the number of tool calling.}
        \label{fig:tool_performance:cum_ac}
    \end{subfigure}
    \hfill 
    \begin{subfigure}{0.48\linewidth}
        \centering
        \includegraphics[width=\linewidth]{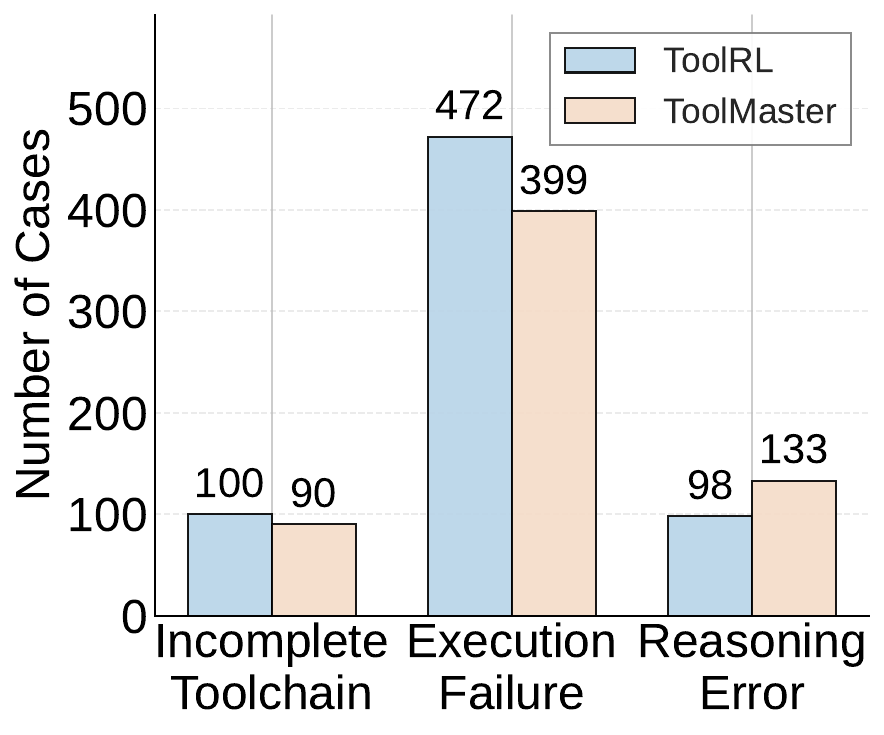}
        \caption{Distribution of tool-calling error types.}
        \label{fig:tool_performance:error-type}
    \end{subfigure}
    \caption{Tool-calling analyses of \method{} in out-of-domain scenarios. We use Qwen2.5-7B-Instruct as the backbone model in experiments and conduct experiments on the ToolHop dataset.}
    \label{fig:tool_performance}
\end{figure}

\subsection{Effectiveness of \method{} in Out-of-Domain Tool-Calling Scenarios}
\label{analyse1}
As shown in Figure~\ref{fig:tool_performance}, we analyze the effectiveness of \method{} in out-of-domain tool-calling scenarios to evaluate its generalization capability and tool-calling behavior.

First, we examine tool trial effectiveness in Figure~\ref{fig:tool_performance:cum_ac} by plotting accumulated answer correctness against the number of tool calling steps. In out-of-domain settings, the model is often unfamiliar with the tools required for the given problems; therefore, conducting appropriate tool calls is crucial for evaluating tool-learning methods. The results show that the correctness of all models increases sharply before the 7-th step, indicating that necessary tool-calling steps are essential for answering the questions. Tool calls typically occur during the execution stage and serve as critical intermediate steps for question answering. As the number of tool calls increases, both Vanilla LLM and ToolRL exhibit plateauing correctness, whereas \method{} continues to improve. This suggests that \method{} is able to conduct effective additional tool trials that better facilitate tool execution to produce accurate results.
As shown in Figure~\ref{fig:tool_performance:error-type}, we further analyze the distribution of tool-calling errors across three categories: Incomplete Toolchain, Execution Failure, and Reasoning Error. These categories are judged using a stronger LLM, DeepSeek-V3.1. The results indicate that \method{} achieves the most notable improvements in the Execution Failure category, demonstrating that tool calling benefits significantly from tool trials, which helps avoid incorrect tool invocations and parameter-passing errors.

\begin{figure}[t]
    \centering
    \begin{subfigure}{0.48\linewidth}
        \centering
        \includegraphics[width=\linewidth]{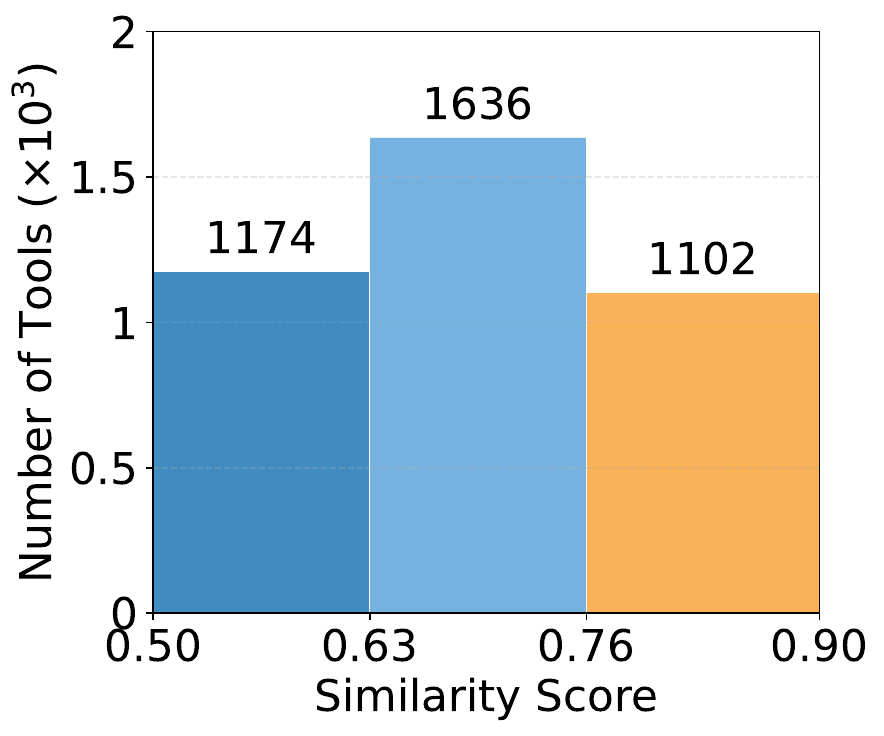}
        \caption{Distribution of tools across different similarity levels.}
        \label{fig:similarity-sub1}
    \end{subfigure}
    \hfill 
    \begin{subfigure}{0.48\linewidth}
        \centering
        \includegraphics[width=\linewidth]{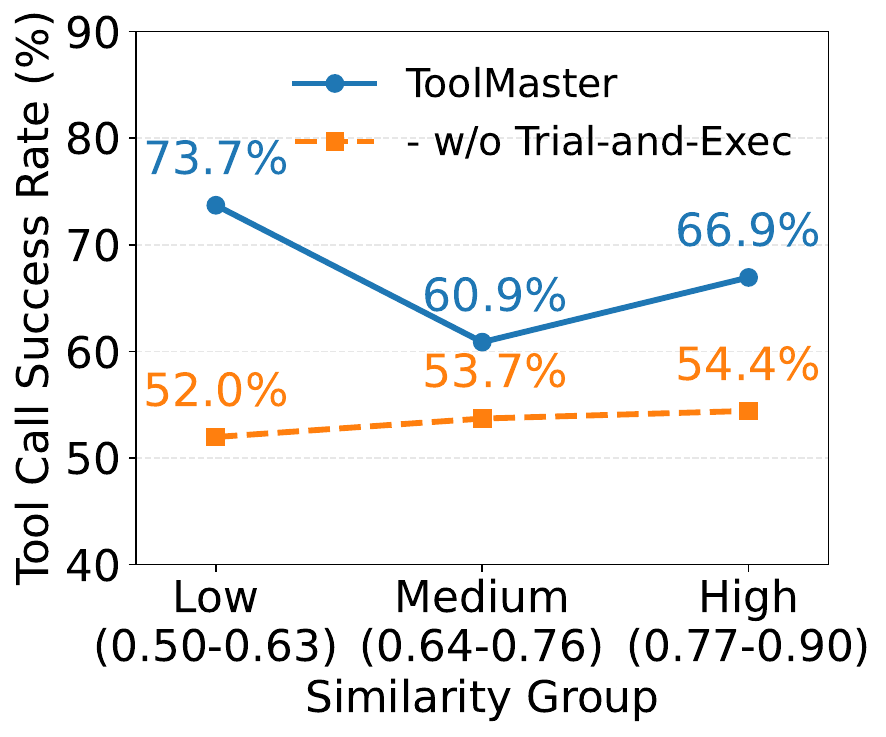}
        \caption{Tool calling success rates for different similarity groups.}
    \label{fig:similarity-sub2}
    \end{subfigure}

    \hfill 
    \begin{subfigure}{0.48\linewidth}
        \centering
        \includegraphics[width=\linewidth]{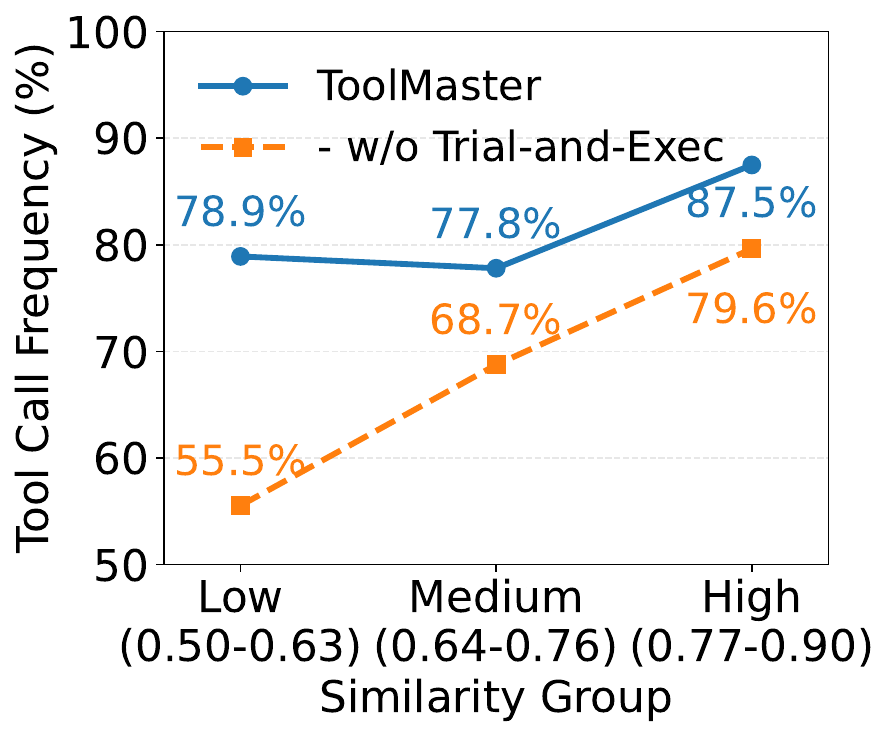}
        \caption{Tool calling frequency of test instances using tools with different similarity levels.}
        \label{fig:similarity-sub3}
    \end{subfigure}
    \hfill 
    \begin{subfigure}{0.48\linewidth}
        \centering
        \includegraphics[width=\linewidth]{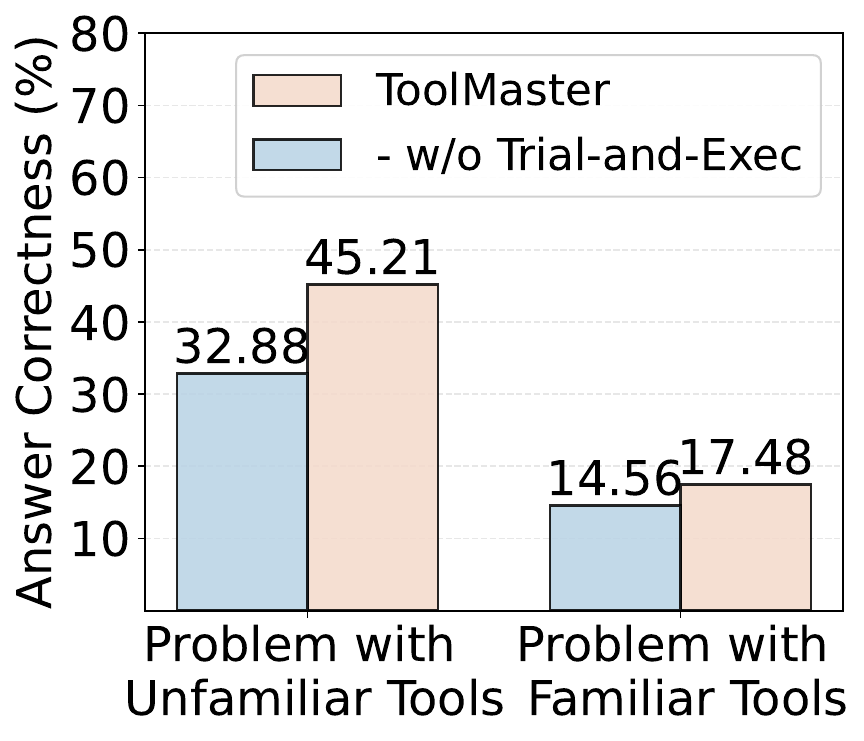}
        \caption{Answer correctness for familiar and unfamiliar tools.}
        \label{fig:similarity-sub4}
    \end{subfigure}

    \caption{Characteristics of \method{} in tool usage under varying degrees of similarity to the training data. This experiment uses Qwen2.5-7B-Instruct as the backbone model and is evaluated on the ToolHop dataset. Figures~\ref{fig:similarity-sub1} and \ref{fig:similarity-sub2} illustrate the distribution of tools and their calling success rates based on similarity between tool documentation and tools observed during training. Figures~\ref{fig:similarity-sub3} and \ref{fig:similarity-sub4} are plotted over test instances, categorized by the tool with the lowest similarity score in the gold tool set.}
    \label{fig:smilarity}
\end{figure}

\subsection{Tool Use Generalization of \method{}}
\label{sec:memorization_bias}
To evaluate the tool-use generalization capability of \method{}, we adopt \method{} w/o Trial-and-Exec as the baseline, which relies solely on golden tool-use trajectories during optimization.

In this experiment, we further categorize the ground-truth tools used in ToolHop into Low, Medium, and High similarity groups. For similarity computation, we employ the Qwen3-Embedding-8B model~\cite{qwen3embedding} to obtain vector representations of tool documentation, and calculate similarity scores using the dot product. We then evaluate the performance of \method{} in terms of the calling success rate of all golden tools, and tool-calling frequency when tools from all three similarity groups are required. Finally, we show the effectiveness of \method{} in handling both familiar and unfamiliar questions that involve tools with varying similarity levels.

As shown in Figure~\ref{fig:similarity-sub1}, we first present the distribution of tools across different similarity levels. The results indicate that the tools are nearly uniformly distributed among the three groups, highlighting the necessity of evaluating tool use across varying similarity levels. We then report the tool-calling success rates for different similarity groups in Figure~\ref{fig:similarity-sub2}. The evaluation results show that \method{} consistently outperforms the baseline, with particularly notable improvements in the Low and High groups. These findings suggest that the tool trial phase enables the model to accumulate practical tool-use experience, which substantially improves tool-calling success for tools with low similarity. Moreover, the gains observed in the High similarity group indicate that \method{} alleviates overfitting to the golden tool-use trajectories.

Next, we randomly sample 330 test instances that require collaborative use of tools from all three similarity groups to solve the query. We report the tool-calling frequency in Figure~\ref{fig:similarity-sub3}. The results show that the baseline model prefers tools from the High similarity group, revealing an unnecessary tool-calling bias. In contrast, \method{} effectively mitigates this bias and exhibits a nearly uniform calling frequency across all groups. Furthermore, we categorize queries into unfamiliar (requiring at least one tool from the Low similarity group) and familiar (all required tools belong to the High similarity group) based on the golden tool set. As shown in Figure~\ref{fig:similarity-sub4}, the evaluation results show that \method{} yields larger performance gains on problems involving unfamiliar tools, highlighting its strong robustness and generalization ability in handling real-world tool usage scenarios.

\section{Conclusion}
This paper introduces \method{}, a novel framework that applies a Trial-and-Execution paradigm to optimize tool-augmented language models. Specifically, \method{} trains LLMs to imitate teacher trajectories that explicitly incorporate tool trials, where tools are invoked to obtain feedback prior to the execution phase. In addition, an RL stage is employed to further jointly coordinate trial strategies. Extensive experimental results demonstrate that \method{} effectively benefits from tool trial interactions, enabling models to better handle unfamiliar tools. 


\section*{Limitations}


Although \method{} demonstrates its effectiveness in improving the robustness and generalization of tool usage, the efficiency of the inference process is still constrained by the inherent nature of the trial-and-execution paradigm. Specifically, since \method{} relies on generating additional trial steps to proactively verify assumptions, the total inference time is inevitably constrained by the latency of receiving responses from external tools during the trial phase. Additionally, \method{} can be applied to diverse real-world environments containing tools with varying functionalities and shows its effectiveness. The safety of deployment may be compromised when tools that induce side effects (e.g., data modification) are involved, due to the model's autonomous tendency to explore unknown tool behaviors. This further underscores the importance of implementing rigorous safety guardrails or sandboxed environments when deploying \method{} in high-stakes applications.


\bibliography{citation}
\clearpage

\appendix

\section{Appendix}
\subsection{License}
The licenses of the resources used in this study are as follows: StableToolBench is released under the Apache License 2.0; ToolHop is released under the CC BY 4.0 license; and RestBench-TMDB is distributed under the MIT License.

\subsection{Details of Implementation}
\label{sec:Details of Implementation}
This subsection provides additional implementation details for our experimental setup.

\textbf{Data Synthesis Prompt.}
Table~\ref{tab:DataSynthesisPrompt} presents the prompt template used during data synthesis. This prompt is designed to guide the assistant to resolve user queries exclusively through tool usage. It first requires the construction of a global plan, which must explicitly include an ``exploration phase'' for verifying tool functionality using sample inputs before addressing the main task. Subsequently, the prompt enforces a structured execution procedure consisting of sub-goal decomposition, validation, and backtracking. Any intermediate failure must trigger strategy revision. By enforcing explicit reasoning traces and iterative problem-solving, this design enables the synthesis of high-quality interaction data that captures realistic agent behaviors, including exploratory tool testing and error recovery.

\textbf{Trajectory Filter Prompt.}
Table~\ref{tab:TrajectoryFilterPrompt} illustrates the prompt used to filter generated trajectories. This prompt acts as an expert evaluator that rigorously inspects conversation logs to determine whether the LLM exhibits targeted advanced problem-solving behaviors. Specifically, it enforces the presence of three mandatory components: global planning and decomposition, explicit tool exploration, and self-correction. The prompt verifies whether the agent actively attempts to ``test'' or understand tool functionality and whether it demonstrates resilience through self-correction upon encountering errors. By defining precise evaluation criteria, such as accepting narrative descriptions as valid plans while strictly requiring exploratory intent, we ensure that only trajectories containing high-quality autonomous reasoning patterns are retained for training.

\textbf{System Prompt.}
Table~\ref{tab:SystemPrompt} shows the system prompt template applied consistently during both training and inference to ensure behavioral alignment. This prompt standardizes the model's operational protocol by enforcing a structured reasoning workflow: first generating a global plan, then executing steps with explicit reasoning enclosed in \texttt{<think>} tags, and finally invoking tools using a predefined XML format (\texttt{<tool\_call>}). By maintaining an identical prompt structure across stages, we ensure that the model internalizes the correct conventions for tool definitions (\texttt{<tools>}), intermediate reasoning, and final answer generation (\texttt{<answer>}).

\textbf{\method{} Training Details.}
During training, we first apply supervised fine-tuning (SFT) followed by reinforcement learning using Group Relative Policy Optimization (GRPO). SFT is conducted for 3 epochs with a per-device batch size of 1, gradient accumulation steps of 16, a learning rate of \(1 \times 10^{-5}\), and a cosine scheduler with 4\% warmup. The maximum sequence length is set to 8{,}192 tokens. GRPO training uses a learning rate of \(1 \times 10^{-6}\), a per-device batch size of 4, gradient accumulation steps of 2, and 4 generations per prompt. The KL divergence coefficient \(\beta\) is set to 0.002, with 2 iterations per update. The maximum prompt length is 1{,}024 tokens and the maximum completion length is 4{,}096 tokens. All training is performed using mixed precision (bf16) with gradient checkpointing enabled. All experiments are conducted on NVIDIA A800 GPUs. During testing, we fix the temperature to 0.1 and \texttt{max\_tokens} to 8{,}192 across all models to ensure consistent evaluation.

\textbf{Reward Function for GRPO Training.}
The reward function consists of two components: a format reward and a correctness reward. The format reward verifies whether the output adheres to the prescribed protocol, requiring reasoning to be enclosed in \texttt{<think>} tags, tool calls in \texttt{<tool\_call>} tags, and the final answer in \texttt{<answer>} tags. The correctness reward is a binary score (0 or 1) assigned by a capable evaluator model, which determines whether the response fully resolves the query. The evaluation prompt is provided in Table~\ref{tab:passrate_status_prompt}.

\textbf{Error Analysis Prompt.}
To explicitly analyze the distribution of tool-calling errors presented in Section~\ref{analyse1}, we employ a judge model to classify failed trajectories. Table~\ref{tab:error_judge_prompt} presents the prompt template used for this error type analysis. This prompt instructs the evaluator to categorize errors into three distinct types: Under-calling, Tool Execution Failure, and Reasoning Discontinuity, based on the provided taxonomy.

\subsection{Details of Datasets Used in Experiments}
\label{sec:More Details on Experiment}

This section provides detailed descriptions of the three benchmarks used in our evaluation: StableToolBench, TMDB, and ToolHop. These benchmarks are selected to cover a broad spectrum of tool-learning challenges, ranging from API stability and orchestration to complex multi-hop reasoning.

\textbf{StableToolBench.}
StableToolBench is divided into six subsets based on difficulty and instruction type. I1 Instruction (I1 Inst), I1 Category (I1 Cat), and I1 Tool focus on single-tool scenarios grounded in documentation, category-level descriptions, or specific functionalities. I2 Category (I2 Cat) and I2 Instruction (I2 Inst) introduce compositional reasoning involving two distinct tools. Finally, I3 Instruction (I3 Inst) contains complex queries that require coordinating three tools. The primary evaluation metric is the Solvable Pass Rate (SoPR), computed by an evaluator model using the prompt provided in Table~\ref{tab:passrate_status_prompt}. An illustrative example is shown in Table~\ref{tab:stb-case}.

\textbf{TMDB.}
The TMDB dataset~\cite{song2023restgpt} simulates a RESTful API environment for movie-related data and evaluates the ability to navigate complex API schemas. Tasks require answering natural language queries about movies, actors, and ratings (e.g., ``Find the release date of the movie directed by X starring Y''). This benchmark involves chaining over 50 distinct API endpoints, requiring models to perform multi-step operations such as entity identification, detail retrieval, and result filtering while correctly handling interdependent parameters. An illustrative example is shown in Table~\ref{tab:tmdb-case}.

\textbf{ToolHop.}
ToolHop~\cite{ye2025toolhop} emphasizes multi-hop reasoning with more than 3{,}000 locally executable tools, removing network latency while preserving functional complexity. Queries are inherently compositional, where the output of one tool (e.g., currency conversion) becomes the input to subsequent tools. Performance is evaluated based on both final answer accuracy against a gold standard and the validity of the execution path. An illustrative example is shown in Table~\ref{tab:toolhop-case}.

\subsection{More Details of Baseline Models}

To assess the effectiveness of our proposed framework, we compare it against a diverse set of baselines spanning traditional supervised fine-tuning and advanced reinforcement learning paradigms. These baselines represent prominent approaches to tool-use learning, covering both data-centric trajectory distillation and policy-centric optimization strategies.

Distill (SFT) serves as a supervised fine-tuning baseline trained on high-quality tool-use trajectories generated by DeepSeek-V3.1. These trajectories function as gold-standard execution paths, providing direct supervision from user instructions to correct API calls and final answers. This baseline primarily evaluates the model's capacity to imitate expert behavior via next-token prediction.

ToolLLM~\cite{qin2023toolllm} is a representative data-centric framework that emphasizes high-quality instruction-tuning data construction. It employs a Depth-First Search Decision Tree (DFSDT) to explore solution spaces in complex tool-use scenarios, enabling recovery from failed attempts and identification of optimal execution paths. We re-implement ToolLLM on the Qwen2.5-7B base model to ensure a fair and up-to-date comparison.

StepTool~\cite{steptool} focuses on fine-grained optimization of the tool-calling process. Instead of relying solely on sparse end-of-trajectory rewards, StepTool adopts Proximal Policy Optimization (PPO) with step-wise rewards. This design enables explicit credit assignment for correct intermediate tool invocations and facilitates learning dependencies among sequential API calls.

FTRL~\cite{ye2025ftrl} is an RL-based approach built on the GRPO framework that leverages environmental feedback as a primary signal for policy updates. By tracing feedback loops, FTRL improves the model's ability to dynamically adjust trajectories based on execution outcomes. In this work, we re-implement FTRL using our training dataset to facilitate a fair comparison under identical experimental settings.

ToolRL~\cite{qian2025toolrl} also adopts the GRPO framework but differentiates itself through its reward design. It optimizes the trade-off between tool-call accuracy and final answer quality by comparing groups of generated trajectories, encouraging the model to favor execution paths that are both successful and schema-compliant.

\subsection{Case study}
The comparison between Table~\ref{tab:correct_sequence} and Table~\ref{tab:failed_sequence} underscores the critical importance of the Trial Phase. The baseline method (ToolRL) fails due to parameter hallucination (e.g., inventing \texttt{output\_format}) and an improper tool selection strategy, resulting in cascading API errors. In contrast, \method{} leverages the Trial Phase to first verify the functionality of the \texttt{extract\_first\_name} tool. This preliminary exploration validates the tool schema and correctly resolves initial sub-goals (name extraction), thereby enabling accurate and error-free execution of subsequent multi-hop reasoning steps, such as identifying siblings and computing letter differences.

\subsection{Cases of Tools with Different Embedding-based Similarity}
To validate the rationale for using embedding similarity to distinguish between \textit{Familiar} and \textit{Unfamiliar} tools, we conduct a comparative analysis of three representative cases (Tables~\ref{tab:moderate_similarity_case}, \ref{tab:similarity_case}, and \ref{tab:high_similarity_case}). The progression of similarity scores illustrates the model's sensitivity in quantifying functional substitutability.

The low-similarity case (Table~\ref{tab:moderate_similarity_case}, score 0.56) demonstrates the model's ability to distinguish divergent intents (extraction vs.\ generation) despite shared domain keywords, correctly categorizing such tools as \textit{Unfamiliar} to prevent misuse. The medium-similarity case (Table~\ref{tab:similarity_case}, score 0.76) highlights inferential reasoning, where a ``Historical Figures'' endpoint implicitly satisfies a ``Family Relationship'' query and is therefore classified as potentially \textit{Familiar}. The high-similarity case (Table~\ref{tab:high_similarity_case}, score 0.86) confirms that explicit alignment in domain-specific terminology (e.g., ``occupation'', ``title'') yields near-synonymous interpretations. This graded spectrum of semantic alignment—from surface relevance to deep functional equivalence—provides strong empirical support for our classification strategy: high-overlap tools enable experience transfer (Exploitation), whereas lower-overlap tools necessitate new learning (Exploration).

\begin{table}[t]
\centering
\small

\begin{tabular}{lcccc}
\toprule

\textbf{Method} & \textbf{I1} & \textbf{I2} & \textbf{I3} & \textbf{Avg.} \\
\midrule
Vanilla  & 2.30 & 2.62 & 4.13 & 2.71 \\
Distill (SFT)    & 4.61 & 4.59 & 7.87 & 5.15 \\
ToolLLM    & 5.15 & 5.49 & 7.41 & 5.64 \\
ToolRL     & 3.45 & 3.39 & 4.66 & 3.63 \\
FTRL       & 4.35 & 4.12 & 6.56 & 4.64 \\
ToolMaster & 5.27 & 5.16 & 8.59 & 5.78 \\
\bottomrule
\end{tabular}
\caption{Comparison of the average number of tool calls across different methods. All models are implemented using Qwen2.5-7B-Instruct.}
\label{tab:tool call num}
\end{table}

\subsection{Efficiency Analysis of \method{}}
We further analyze inference efficiency by comparing the average number of tool invocations required to complete tasks, as reported in Table~\ref{tab:tool call num}.

Although simpler baselines such as Zero-shot and ToolRL exhibit lower average tool usage, this typically reflects an inability to sustain the multi-step reasoning chains required for complex queries rather than genuine efficiency. Importantly, compared with the competitive baseline ToolLLM, ToolMaster exhibits a highly comparable tool usage profile with only a marginal increase in calls. This indicates that ToolMaster introduces minimal computational overhead, and the slight increase in inference cost is a reasonable trade-off for the substantial gains in task success rate and reasoning robustness.

\begin{table}[t]
    \centering
    \small
    \begin{tabular}{l c c c}
        \toprule
        \textbf{Method} & \textbf{TMDB} & \textbf{ToolHop} & \textbf{Avg.} \\
        \midrule
        Vanilla & 72.91 & 66.36 & 69.64 \\
        Distill (SFT) & 80.17 & 72.56 & 76.37 \\
        ToolLLM & 78.58 & \underline{75.45} & \underline{77.02} \\
        FTRL & 73.92 & 61.21 & 67.57 \\
        ToolRL & \underline{83.67} & 63.20 & 73.44 \\
        ToolMaster & \textbf{91.25} & \textbf{80.84} & \textbf{86.05} \\
        \bottomrule
    \end{tabular}
    \caption{Comparison of Correct Path Rates on TMDB and ToolHop benchmarks. All models are implemented using Qwen2.5-7B-Instruct.}
    \label{tab:correct_path_rate}
\end{table}

\subsection{Tool Selection Analysis}
To further assess the fidelity of intermediate reasoning, we conduct an additional evaluation focusing on tool selection correctness. We employ the Correct Path Rate metric on both the TMDB and ToolHop benchmarks. This metric is defined as the recall of the tool usage trajectory, measuring the proportion of ground-truth tools correctly invoked by the model relative to the total set required to solve the query.

As shown in Table~\ref{tab:correct_path_rate}, ToolMaster consistently outperforms all baseline methods in tool selection accuracy. While baseline models frequently fail to identify the complete set of required tools for complex queries, our method aligns more closely with the ground-truth execution paths. These results suggest that the trial-and-execution paradigm effectively guides the model to eliminate irrelevant options and accurately identify appropriate tools, thereby establishing a stronger foundation for correct final answers.

\begin{table*}[t]
\centering
\small
\setlength{\tabcolsep}{6pt}
\renewcommand{\arraystretch}{1.08}

\begin{tabular}{|p{0.985\textwidth}|}
\hline
\textbf{Case: StableToolBench} \\
\hline

\begin{minipage}[t]{\linewidth}
\textbf{Query ID:} 7257

\vspace{0.5ex}
\textbf{Question:}\\
I'm a blogger and I want to verify the email addresses of my subscribers. Can you validate the emails of my subscribers using the Email Validate Regex API? Additionally, fetch the inbox messages for the email address
\texttt{p1amvpvxfh@bestparadize.com}
using the Read Inbox API to check for any collaboration opportunities or feedback from my readers.

\vspace{0.5ex}
\textbf{API List (candidate tools):}
\begin{itemize}[leftmargin=*, itemsep=0ex, topsep=0ex]
\item \texttt{email\_validator\_v3\_email\_validate\_regex} (Email Validate Regex).\\
Required params: \texttt{email} (STRING), default \texttt{test@gmail.com}.
\item \texttt{temp\_mail\_read\_inbox} (Read inbox an email).\\
Required params: \texttt{email} (string).
\item \texttt{account\_verifyer\_instagram\_account\_verifyer} (to verify Instagram account).\\
Required params: none.
\item \texttt{emails\_verifier\_verify\_email} (Allows verifying email addresses. Checks if emails are deliverable.)\\
Required params: \texttt{query} (STRING), default \texttt{support@outscraper.com}.
\end{itemize}

\vspace{0ex}
\textbf{Data characteristics:}
\begin{itemize}[leftmargin=*, itemsep=0ex, topsep=0ex]
\item Multi-intent request: email validation + inbox retrieval in one query.
\item Tool redundancy: multiple email-verification APIs with different parameter names (\texttt{email} vs.\ \texttt{query}).
\item Noisy candidate tool: an Instagram verification API appears unrelated to the user's request.
\end{itemize}
\end{minipage}
\\ \hline
\end{tabular}

\caption{An example case from StableToolBench.}
\label{tab:stb-case}
\end{table*}

\begin{table*}[t]
\centering
\small
\setlength{\tabcolsep}{6pt}
\renewcommand{\arraystretch}{1.08}

\begin{tabular}{|p{0.985\textwidth}|}
\hline
\textbf{Case: ToolHop} \\
\hline

\begin{minipage}[t]{\linewidth}
\textbf{Query ID:} 993

\vspace{0.5ex}
\textbf{Question:}\\
The submission deadline is at 08 January 2008, 09:37, Anywhere on Earth (AoE).
At what date and time is the deadline in the county in which Kimbrough Memorial Stadium is located?

\vspace{0.5ex}
\textbf{API List (candidate tools):}
\begin{itemize}[leftmargin=*, itemsep=0ex, topsep=0ex]
\item \texttt{geo\_entity\_locator}
\item \texttt{geo\_locator}
\item \texttt{geo\_time\_zone\_finder}
\item \texttt{timezone\_difference\_calculator}
\item \texttt{advanced\_timezone\_converter}
\end{itemize}

\vspace{0ex}
\textbf{Data characteristics:}
\begin{itemize}[leftmargin=*, itemsep=0ex, topsep=0ex]
\item Multi-hop reasoning with intermediate entities: stadium $\rightarrow$ city $\rightarrow$ county $\rightarrow$ timezone $\rightarrow$ conversion.
\item The execution path is compositional and inspectable: each intermediate output serves as the input for the next tool.
\end{itemize}
\end{minipage}
\\ \hline
\end{tabular}

\caption{An example case from ToolHop.}
\label{tab:toolhop-case}
\end{table*}


\begin{table*}[t]
\centering
\small
\setlength{\tabcolsep}{6pt}
\renewcommand{\arraystretch}{1.08}

\begin{tabular}{|p{0.985\textwidth}|}
\hline
\textbf{Case: TMDB} \\
\hline

\begin{minipage}[t]{\linewidth}
\textbf{Query ID:} q3

\vspace{0.5ex}
\textbf{Question:}\\
give me a image for the collection Star Wars

\vspace{0.5ex}
\textbf{API List (candidate tools):}
\begin{itemize}[leftmargin=*, itemsep=0.2ex, topsep=0.2ex]
\item \texttt{GET\_search\_collection}
\item \texttt{GET\_collection\_collection\_id\_images}
\item \texttt{GET\_tv\_tv\_id\_recommendations}
\item \texttt{GET\_movie\_latest}
\item \texttt{GET\_tv\_tv\_id\_season\_season\_number}
\item \texttt{GET\_discover\_tv}
\item \texttt{GET\_search\_tv}
\item \texttt{GET\_tv\_tv\_id\_similar}
\item \texttt{GET\_tv\_tv\_id\_images}
\item \texttt{GET\_review\_review\_id}
\item \texttt{GET\_tv\_on\_the\_air}
\item \texttt{GET\_movie\_movie\_id\_release\_dates}
\item \texttt{GET\_movie\_movie\_id\_reviews}
\item \texttt{GET\_tv\_airing\_today}
\item \texttt{GET\_tv\_tv\_id\_credits}
\item \texttt{GET\_movie\_top\_rated}
\item \texttt{GET\_discover\_movie}
\item \texttt{GET\_person\_person\_id\_tv\_credits}
\item \texttt{GET\_tv\_top\_rated}
\item \texttt{GET\_movie\_movie\_id\_credits}
\end{itemize}

\vspace{0ex}
\textbf{Data characteristics:}
\begin{itemize}[leftmargin=*, itemsep=0ex, topsep=0ex]
\item Two-hop tool chaining: search (name $\rightarrow$ id) then images (id $\rightarrow$ image paths).
\item Heavy tool-list noise: many TV/movie endpoints are irrelevant to a \emph{collection image} request.
\end{itemize}
\end{minipage}
\\ \hline
\end{tabular}

\caption{An example case from TMDB.}
\label{tab:tmdb-case}
\end{table*}

\begin{table*}[h]
\centering
\renewcommand{\arraystretch}{1.3}

\small
\begin{tabular}{|p{\textwidth}|}
\hline
\textbf{Input Details} \\ \hline
\textbf{Query:} How many additional letters does the first name of Lori Lyons have compared to the first name of Frances Freeling Broderip's sibling?

\textbf{Available APIs:} \\
\texttt{family\_relationship\_finder(individual\_name: str, relationship\_type: str, ...)} \\
\texttt{enhanced\_family\_relationship\_finder(person\_name: str, data\_source: str, ...)} \\
\texttt{extract\_first\_name(names: array)} \\
\texttt{count\_letters(input: str)} \\
\texttt{advanced\_arithmetic\_calculator(operation: str, numbers: array)} \\

\hline
\textbf{Step 1: Trial Phase} \\ \hline

\textbf{Thought:} <think>1. Exploration Phase: First, I need to understand the functionality of the tools. I'll start by exploring the `extract\_first\_name` tool to extract the first names from the given names. I'll test with "Lori Lyons" and "Frances Freeling Broderip" to see how the tool works.\\2. Sub-goal 1: Extract the first name "Lori" from "Lori Lyons"...
\\I'll start with the exploration phase by testing the `extract\_first\_name` tool on "Lori Lyons" to see how it works.</think>\\

\textbf{Action:} <tool\_call>{"name": "extract\_first\_name", "arguments": {"full\_names": ["Lori Lyons"]}}</tool\_call>" 

\textbf{Observation:} 
""Lori""
 \\ 
\hline

\textbf{Step 2: Trial Phase} \\ \hline

\textbf{Thought:}Great! The tool correctly extracted "Lori" from "Lori Lyons". Now, I'll test it on "Frances Freeling Broderip" to extract "Frances".</think>

\textbf{Action:} <tool\_call>{"name": "extract\_first\_name", "arguments": {"full\_names": ["Frances Freeling Broderip"]}}</tool\_call>"

\textbf{Observation:} Frances
 \\ 
\hline

\textbf{Step 3: Execution Phase} \\ \hline

\textbf{Thought:} <think>Perfect! The tool works well for both names. Now, I need to find the siblings of "Frances Freeling Broderip" using the `family\_relationship\_finder` tool. I'll set the relationship type to "sibling" and use "Frances Freeling Broderip" as the individual.</think>\\
\textbf{Action:} <tool\_call>{"name": "family\_relationship\_finder", "arguments": {"individual\_name": "Frances Freeling Broderip", "relationship\_type": "sibling"}}</tool\_call> \\
\textbf{Observation:} Tom Hood
 \\ 
\hline

\textbf{Step 4: Execution Phase} \\ \hline

\textbf{Thought:}  "<think>Interesting! The tool found one sibling for "Frances Freeling Broderip": "Tom Hood". Now, I need to extract the first name "Tom" from "Tom Hood" using the extract\_first\_name tool.</think>
\\

\textbf{Action:} <tool\_call>{"name": "extract\_first\_name", "arguments": {"full\_names": ["Tom Hood"]}}</tool\_call> 

\textbf{Observation:} Tom
 \\ 
\hline

\textbf{Step 5: Execution Phase} \\ \hline

\textbf{Thought:}  "<think>Excellent! The tool extracted "Tom" from "Tom Hood". Now, I have both first names: "Lori" (from "Lori Lyons") and "Tom" (from "Tom Hood"). I'll use the `count\_letters` tool to count the letters in each name. I'll start with "Lori".</think>

\textbf{Action:} <tool\_call>{"name": "count\_letters", "arguments": {"input": "Lori"}}</tool\_call>" 

\textbf{Observation:} 4\\ 
\hline
\textbf{Step 6: Execution Phase} \\ \hline
\textbf{Thought:} <think>Great! "Lori" has 4 letters. Now, I'll count the letters in "Tom".</think>

\textbf{Action:} <tool\_call>{"name": "count\_letters", "arguments": {"input": "Tom"}}</tool\_call> \\

\textbf{Observation:} 3\\
\hline
\textbf{Step 7: Final Response} \\ \hline
 \textbf{Thought:}<think>Perfect! "Tom" has 3 letters. The final step is to calculate the difference: 4 (Lori) - 3 (Tom) = 1. So, "Lori" has 1 more letter than "Tom". I'll validate all steps: we successfully extracted the first names, found the sibling, and counted the letters. The answer is 1.</think>\\
 \textbf{Final Answer:} <answer>The first name "Lori" has 1 additional letter compared to the first name "Tom" of Frances Freeling Broderip's sibling.</answer> \\

\hline
\end{tabular}
\caption{Case study 1: The correct solution path generated by \method{}.}
\label{tab:correct_sequence}
\end{table*}

\begin{table*}[h]
\centering
\renewcommand{\arraystretch}{1.3}

\small
\begin{tabular}{|p{\textwidth}|}
\hline
\textbf{Input Details} \\ \hline
\textbf{Query:} How many additional letters does the first name of Lori Lyons have compared to the first name of Frances Freeling Broderip's sibling?

\textbf{Available APIs:} \\
\texttt{family\_relationship\_finder(individual\_name: str, relationship\_type: str, ...)} \\
\texttt{enhanced\_family\_relationship\_finder(person\_name: str, data\_source: str, ...)} \\
\texttt{extract\_first\_name(names: array)} \\
\texttt{count\_letters(input: str)} \\
\texttt{advanced\_arithmetic\_calculator(operation: str, numbers: array)} \\
\hline

\textbf{Step 1: Execution Phase} \\ \hline
\textbf{Thought:} <think>I need to find the first names of Lori Lyons and Frances Freeling Broderip's sibling using the family relationship finder tools. First, I'll start by finding the first name of Lori Lyons using the family relationship finder.</think> \\
\textbf{Action:} <tool\_call>\{"name": "family\_relationship\_finder", "arguments": \{"individual\_name": "Lori Lyons", "relationship\_type": "child", "data\_source": "FamilySearch", "output\_format": "json"\}\}</tool\_call> \\
\textbf{Observation:} "an error occured when call family\_relationship\_finder: family\_relationship\_finder() got an unexpected keyword argument 'output\_format'" \\
\hline

\textbf{Step 2: Execution Phase} \\ \hline
\textbf{Thought:} <think>I received an error because I specified "output\_format" which is not a required or optional parameter for the family\_relationship\_finder. I'll try without it.</think> \\
\textbf{Action:} <tool\_call>\{"name": "family\_relationship\_finder", "arguments": \{"individual\_name": "Lori Lyons", "relationship\_type": "child", "data\_source": "FamilySearch"\}\}</tool\_call> \\
\textbf{Observation:} "Error: No data found for Lori Lyons with relationship type child." \\
\hline

\textbf{Step 3: Execution Phase} \\ \hline
\textbf{Thought:} <think>I got an error for Lori Lyons. Let me try with a different relationship type, like "parent", to see if that works better.</think> \\
\textbf{Action:} <tool\_call>\{"name": "family\_relationship\_finder", "arguments": \{"individual\_name": "Lori Lyons", "relationship\_type": "parent", "data\_source": "FamilySearch"\}\}</tool\_call> \\
\textbf{Observation:} "Error: 'relationship\_type' must be one of ['father', 'mother', 'sibling', 'child', 'grandparent', 'aunt', 'uncle', 'cousin']." \\
\hline

\textbf{Step 4: Execution Phase} \\ \hline
\textbf{Thought:} <think>I see that "parent" is not a valid relationship type for this tool. The valid types are listed in the enum. Since "child" and "parent" are not working, I'll try "sibling" as it might be relevant for finding first names. ... I'll use the enhanced family relationship finder for this, as it might handle the dataset aspect better.</think> \\
\textbf{Action:} <tool\_call>\{"name": "enhanced\_family\_relationship\_finder", "arguments": \{"person\_name": "Frances Freeling Broderip", "data\_source": "FamilySearch", "relationship\_type": "sibling", "output\_format": "json"\}\}</tool\_call> \\
\textbf{Observation:} "an error occured when call enhanced\_family\_relationship\_finder: enhanced\_family\_relationship\_finder() got an unexpected keyword argument 'output\_format'" \\
\hline

\textbf{Step 5: Execution Phase} \\ \hline
\textbf{Thought:} <think>I received an error for the enhanced family relationship finder as well, due to the "output\_format" parameter. I'll try without it.</think> \\
\textbf{Action:} <tool\_call>\{"name": "enhanced\_family\_relationship\_finder", "arguments": \{"person\_name": "Frances Freeling Broderip", "data\_source": "FamilySearch", "relationship\_type": "sibling"\}\}</tool\_call> \\
\textbf{Observation:} \{"error": "Person not found in the data source."\} \\
\hline

\textbf{Step 6: Final Response} \\ \hline
\textbf{Thought:} <think>I have attempted to retrieve the first names using the available tools, but both queries resulted in errors or data not found. ... Since I've tried the available approaches and encountered limitations, I should provide a final answer based on the findings.</think> \\
\textbf{Final Answer:} <answer>I attempted to retrieve the first names using the available tools, but encountered several limitations. The family relationship finder for "Lori Lyons" resulted in an error when specifying an "output\_format," and when I removed it, it returned "No data found for Lori Lyons with relationship type child." The enhanced family relationship finder for "Frances Freeling Broderip" also had an issue with the "sibling" relationship type, returning an error. Based on the tool descriptions, these tools are designed for finding familial relationships rather than extracting first names from names. ... Due to these limitations, I was unable to retrieve the specific first names for the calculation.</answer> \\
\hline
\end{tabular}
\caption{Case study 1: The failed solution path generated by ToolRL.}
\label{tab:failed_sequence}
\end{table*}

\begin{table*}[h]
\centering
\renewcommand{\arraystretch}{1.3}
\small
\begin{tabular}{|p{\textwidth}|}
\hline
\textbf{Tool Pair Comparison (Calculated Similarity: 0.5616)} \\ \hline

\textbf{Tool A:} \texttt{extract\_first\_letter} \\
\textbf{Description:} An enhanced tool designed to extract the first letter from a given string. It supports various string types and includes options for handling edge cases. \\
\textbf{Parameters:}
\begin{itemize}
    \item \textbf{Required:} \texttt{input} (the source string)
    \item \textbf{Optional:} \texttt{ignore\_whitespace}, \texttt{ignore\_non\_alphabetic}, \texttt{character\_set}, \texttt{return\_format}
\end{itemize} \\
\hline

\textbf{Tool B:} \texttt{random\_word\_api\_get\_word\_by\_start\_and\_contain} \\
\textbf{Description:} Returns a random word that starts with a certain string and contains a certain string (e.g., starts with "fru", contains "le"). \\
\textbf{Parameters:}
\begin{itemize}
    \item \textbf{Required:} \texttt{start} (prefix string), \texttt{substring} (contained string)
\end{itemize}
\textbf{Template Response:} \texttt{word} (string) \\
\hline

\textbf{Analysis:} \\
The model assigns a moderate similarity score (\textbf{0.5616}), reflecting a nuanced understanding of the functional overlap without over-matching.
\begin{itemize}
    \item \textbf{Shared Context:} Both tools operate heavily within the domain of \textit{string manipulation} and \textit{character positioning}. Tool A focuses on the "first letter" (a specific position), while Tool B filters words based on how they "start" (a positional constraint).
    \item \textbf{Distinct Intent:} The score is in low similarity groups because the core intents differ: Tool A is an \textit{extraction} utility (deterministic processing), whereas Tool B is a \textit{generation/retrieval} API (randomized output). The embedding correctly identifies them as related "word/string tools" but distinct enough to avoid confusion in high-precision retrieval tasks.
\end{itemize} \\
\hline
\end{tabular}
\caption{Case study 2: Analysis of low semantic similarity tools.}
\label{tab:moderate_similarity_case}
\end{table*}

\begin{table*}[h]
\centering
\renewcommand{\arraystretch}{1.3}
\small
\begin{tabular}{|p{\textwidth}|}
\hline
\textbf{Tool Pair Comparison (Calculated Similarity: 0.7586)} \\ \hline

\textbf{Tool A:} \texttt{family\_relationship\_finder} \\
\textbf{Description:} An advanced tool designed to identify a variety of family relationships, including parentage, siblings, and offspring, for historical and contemporary figures. \\
\textbf{Parameters:}
\begin{itemize}
    \item \textbf{Required:} \texttt{person\_name}, \texttt{relationship\_type} (enum: father, mother, sibling, child)
    \item \textbf{Optional:} \texttt{historical\_context}, \texttt{data\_source\_preference}, \texttt{date\_range}
\end{itemize} \\
\hline

\textbf{Tool B:} \texttt{historical\_figures\_by\_api\_ninjas\_v1\_historicalfigures} \\
\textbf{Description:} API Ninjas Historical Figures API endpoint. Returns a list of up to 10 people that match the search parameters. \\
\textbf{Parameters:}
\begin{itemize}
    \item \textbf{Optional:} \texttt{name} (e.g., "julius caesar"), \texttt{offset}
\end{itemize}
\textbf{Key Response Fields:} \texttt{parents}, \texttt{children}, \texttt{spouses}, \texttt{born}, \texttt{died} \\
\hline

\textbf{Analysis:} \\
Despite differences in naming conventions and parameter structures, the model assigns a high similarity score (\textbf{0.7586}).
\begin{itemize}
    \item \textbf{Parameter Mapping:} The model detects semantic equivalence between Tool A's \texttt{person\_name} and Tool B's \texttt{name}. Although Tool A requires specific relationship types as inputs, Tool B implicitly covers these relationships in its output schema.
    \item \textbf{Semantic Alignment:} The embedding successfully bridges the gap between a function-oriented tool ("Finder") and a data-oriented endpoint ("Historical Figures"). It understands that retrieving a historical figure's profile (Tool B) is semantically aligned with querying their family tree (Tool A), as evidenced by the shared concepts of "parents" and "children" in their definitions.
\end{itemize} \\
\hline
\end{tabular}
\caption{Case study 3: Analysis of medium semantic similarity tools.}
\label{tab:similarity_case}
\end{table*}

\begin{table*}[h]
\centering
\renewcommand{\arraystretch}{1.3}
\small
\begin{tabular}{|p{\textwidth}|}
\hline
\textbf{Tool Pair Comparison (Calculated Similarity: 0.8551)} \\ \hline

\textbf{Tool A:} \texttt{genealogy\_lookup} \\
\textbf{Description:} An advanced tool for retrieving detailed genealogical information about \textit{historical figures}. Offers functionalities to explore family trees and manage titles. \\
\textbf{Parameters:}
\begin{itemize}
    \item \textbf{Required:} \texttt{name}, \texttt{relationship} (enum: father, mother, spouse, etc.)
    \item \textbf{Optional:} \texttt{title}, \texttt{occupation}, \texttt{historical\_context}, \texttt{geographical\_location}
\end{itemize} \\
\hline

\textbf{Tool B:} \texttt{historical\_figures\_by\_api\_ninjas\_v1\_historicalfigures} \\
\textbf{Description:} API Ninjas Historical Figures API endpoint. Returns a list of people matching search parameters. \\
\textbf{Parameters:}
\begin{itemize}
    \item \textbf{Optional:} \texttt{name}, \texttt{offset}
\end{itemize}
\textbf{Key Response Fields:} \texttt{title}, \texttt{occupation}, \texttt{parents}, \texttt{children}, \texttt{spouses}, \texttt{awards} \\
\hline

\textbf{Analysis:} \\
The model assigns a very high similarity score (\textbf{0.8551}), indicating a near-synonymous functional relationship.
\begin{itemize}
    \item \textbf{Strong Domain Identity:} Unlike the previous case, Tool A explicitly includes "historical figures" in its description, creating a direct lexical and semantic match with Tool B's name and domain.
    \item \textbf{Comprehensive Field Mapping:} The embedding detects a dense alignment between Tool A's input parameters and Tool B's output schema.
    \begin{itemize}
        \item Tool A's optional inputs \texttt{title} and \texttt{occupation} directly map to Tool B's response fields \texttt{title} and \texttt{occupation}.
        \item Tool A's \texttt{relationship} parameter (father, spouse) perfectly corresponds to Tool B's structure (\texttt{parents}, \texttt{spouses}).
    \end{itemize}
\end{itemize} \\
\hline
\end{tabular}
\caption{Case study 4: Analysis of high semantic similarity tools.}
\label{tab:high_similarity_case}
\end{table*}

\begin{table*}[t]
\centering
\setlength{\tabcolsep}{6pt}
\renewcommand{\arraystretch}{1.08}

\begin{tabularx}{\textwidth}{|>{\raggedright\arraybackslash}X|}
\hline
\textbf{Data Synthesis Prompt} \\
\hline

\textbf{Task Description}\\
You are a smart assistant capable of utilizing provided tools to answer users' questions. Your primary strategy should be to use the available tools rather than relying on internal knowledge. Follow this enhanced process to solve problems, incorporating the behaviors of backtracking, setting sub-goals, validation, and exploration.

\begin{enumerate}[leftmargin=*, itemsep=0ex, topsep=0ex]
\item Create a global plan for the query. This plan must include: an exploration phase to understand tool functionality by making initial tool calls with sample inputs. Decomposition of the task into sub-goals when necessary.
\item Execute each step in the plan:
\begin{itemize}[leftmargin=*, itemsep=0ex, topsep=0ex]
\item Record your thought process, formatted as \texttt{<thought></thought>}.
\item Call the appropriate tool by providing its name and parameters in JSON format, like
\texttt{<|tool calls begin|><|tool call begin|>tool's name<|tool sep|>\{\{"parameters1":"value1","parameters2":"value2"\}\}<|tool call end|><|tool calls end|>},
formatted as \texttt{<|tool calls begin|><|tool call begin|><|tool call end|><|tool calls end|>}.
\item Every turn must include one tool call and do not combine multiple tool calls in one turn.
\item Validate whether the result meets the requirements of the current step. If not, backtrack by revising the plan and repeating the relevant steps.
\item If a tool is unavailable or returns an error, consider using alternative tools that can achieve similar results, and adjust the plan accordingly.
\item Try your best to use the tools to obtain information and you can call the tools multiple times if necessary.
\end{itemize}
\end{enumerate}

\textbf{Stopping Criteria}\\
You can stop the problem-solving process when:
\begin{itemize}[leftmargin=*, itemsep=0ex, topsep=0ex]
\item You have obtained a result that fully satisfies the user's request.
\item You have validated that the result meets all requirements.
\item All sub-goals have been successfully addressed.
\end{itemize}

\textbf{Tool Usage Policy}\\
\begin{itemize}[leftmargin=*, itemsep=0ex, topsep=0ex]
\item Always prioritize using the available tools to obtain information.
\item When uncertain about the accuracy or sufficiency of information from tools, perform additional tool calls or validation steps.
\item Do not use internal knowledge and only rely on tools to get information.
\item If a tool is unavailable, look for alternative tools within the provided documentation that can serve as substitutes to achieve the same objective.
\item Include an exploration phase in your plan to better understand tool behavior before applying them to the task. This phase is mandatory and must be completed before proceeding with the main task.
\end{itemize}

\textbf{Tool Documentation}\\
Below is the documentation for available tools: \{tool doc\}.\\
\hline
\end{tabularx}

\caption{Data synthesis prompt template.}
\label{tab:DataSynthesisPrompt}
\end{table*}

\begin{table*}[t]
\centering
\setlength{\tabcolsep}{6pt}
\renewcommand{\arraystretch}{1.08}

\begin{tabularx}{\textwidth}{|>{\raggedright\arraybackslash}X|}
\hline
\textbf{Trajectory Filter Prompt} \\
\hline

You are an expert evaluator of AI Agent reasoning and tool usage. Your task is to analyze a conversation log between a User and an AI Assistant to determine if the Assistant exhibits a specific set of advanced problem-solving behaviors.

You must look for the presence of three distinct behaviors. The Assistant does not need to use exact keywords (like ``Global Plan'' or ``Backtracking''), but the reasoning process in the \texttt{<think>} tags must clearly demonstrate these actions took place.

\textbf{The Three Required Behaviors:}

\begin{enumerate}[leftmargin=*, itemsep=0ex, topsep=0ex]
\item \textbf{Global Planning \& Decomposition:}

The Assistant must set a high-level strategy at the beginning.

It should break complex user queries into smaller, manageable sub-goals or steps.

Criteria: Does the Assistant explicitly map out what it intends to do before jumping into tool calls?

\item \textbf{Tool Exploration (Mandatory):}

The Assistant must demonstrate an intent to ``understand'' or ``test'' a tool before fully relying on it for the final answer.

This can appear as:
\begin{itemize}[leftmargin=*, itemsep=0ex, topsep=0ex]
\item Calling a tool to see its output format (schema exploration).
\item Calling a tool with sample data to verify behavior.
\item Explicitly stating in the thought process that a call is being made to ``explore,'' ``check capabilities,'' or ``understand the response'' (even if using real user data).
\end{itemize}

Criteria: Is there a step where the Agent tries to learn about the tool's behavior rather than just assuming it works perfectly immediately?

\item \textbf{Validation \& Backtracking (Self-Correction):}

Validation: After receiving a tool output, the Assistant must evaluate if the data satisfy the user's request.

Resilience/Backtracking:
\begin{itemize}[leftmargin=*, itemsep=0ex, topsep=0ex]
\item If an error occurs: The Assistant must acknowledge the error and propose a fix, a retry with different parameters, or a substitute tool.
\item If successful: The Assistant validates the data is correct. (Note: If the tool works perfectly, ``backtracking'' is not required, only validation is required).
\end{itemize}

Criteria: Does the Agent verify the results? If things go wrong, does it try to fix them instead of giving up or ignoring the error?
\end{enumerate}

\textbf{Evaluation Rules:}

\begin{itemize}[leftmargin=*, itemsep=0ex, topsep=0ex]
\item Be Lenient on Format: Do not demand specific XML tags or numbered lists for the plan. Narrative paragraphs are acceptable if the logic is present.
\item Contextual Exploration: ``Exploration'' is valid even if the agent uses the user's actual input, provided the intent described in the thought process is to verify how the tool functions or returns data.
\item Partial Trajectories: If the log ends abruptly (e.g., during a retry), judge based on the behaviors exhibited so far. If the agent demonstrated the intent to fix an error, that counts as satisfying the Validation/Backtracking requirement.
\end{itemize}

\textbf{Output Format:}

\begin{itemize}[leftmargin=*, itemsep=0ex, topsep=0ex]
\item Analysis: Briefly describe where you found evidence (or lack thereof) for each of the three behaviors.
\item Result: Output only True if ALL three behaviors are present. Output False if ANY of the three are missing.
\end{itemize}

\\ \hline
\end{tabularx}

\caption{Trajectory filter prompt template.}
\label{tab:TrajectoryFilterPrompt}
\end{table*}

\begin{table*}[t]
\centering
\renewcommand{\arraystretch}{1.15}
\small
\setlength{\tabcolsep}{6pt}

\begin{tabularx}{\textwidth}{|>{\raggedright\arraybackslash}X|}
\hline
\textbf{System Prompt} \\
\hline

You are a smart assistant capable of utilizing provided tools to answer users' questions. Follow this process to solve problems:

\begin{enumerate}[leftmargin=*, itemsep=0.3ex, topsep=0.3ex]
\item Create a global plan for the query.
\item Execute each step in the plan:
\begin{itemize}[leftmargin=*, itemsep=0.3ex, topsep=0.3ex]
\item Record your thought process, formatted as \texttt{<think></think>}.
\item Call the appropriate tool by providing its name and parameters in JSON format.
For each function call, return a json object with function name and arguments within
\texttt{<tool\_call></tool\_call>} XML tags:
\texttt{<tool\_call>\{\{"name": <function-name>, "arguments": <args-json-object>\}\}</tool\_call>}.
\item Get the result returned by the tool, formatted as \texttt{<tool\_response></tool\_response>}.
\end{itemize}
\item After completing all steps, provide the final answer, formatted as \texttt{<answer></answer>}.
\end{enumerate}

You are provided with function signatures within \texttt{<tools></tools>} XML tags:\\
\texttt{<tools>}\\
\{tool-docs\}\\
\texttt{</tools>}\\

\hline
\end{tabularx}

\caption{System prompt template.}
\label{tab:SystemPrompt}
\end{table*}

\begin{table*}[h]
\centering
\renewcommand{\arraystretch}{1.3}

\small
\begin{tabular}{|p{\textwidth}|}
\hline
\textbf{Pass Rate Evaluation Prompt} \\ \hline
You are an assistant responsible for evaluating whether an LLM agent's response should be counted as Pass, Fail, or Unsure in passrate calculations. Your evaluation must consider both the final answer and the complete execution chain. \\
\textbf{Status Determination Rules:} \\
\textbf{Pass:} Answer sufficiently solves query; Execution chain shows successful API calls; Initial errors were corrected; Information verifiable through API responses \\
\textbf{Fail:} API observations show execution errors; Answer contradicts evidence; Information incorrect/invalid; Solution misses core requirements \\
\textbf{Unsure:} Cannot verify authenticity; Insufficient validation data; Need complete reasoning process; \\
\textbf{Output Format:} \\
\begin{minipage}{\textwidth}
\begin{verbatim}
{"content": "Evaluation reasoning", "answer_status": "Pass/Fail/Unsure"}
\end{verbatim}
\end{minipage} \\
\textbf{Required Input:} Original query; Final answer; Complete execution chain with API responses \\ [0.5em]
\hline
\end{tabular}
\caption{Pass Rate evaluation prompt template followed by ToolMVR~\cite{ToolMVR}.}
\label{tab:passrate_status_prompt}
\end{table*}
\begin{table*}[h]
\centering
\small
\renewcommand{\arraystretch}{1.2}
\begin{tabular}{|p{0.96\linewidth}|}
\hline
\textbf{Error Type Judge Prompt template} \\
\hline
You are an expert analyst of Large Language Model (LLM) Agent behaviors. Your task is to analyze failed execution trajectories and classify the \textbf{PRIMARY} error type into EXACTLY one of the following 3 categories. \\
\\
\textbf{Input Data} - `ground\_truth': The correct answer. - `response': The trajectory of thoughts, tool calls, and tool outputs. \\
\textbf{Classification Taxonomy} Please identify the \textbf{Root Cause} or the most \textbf{Fatal Error} that led to the failure. Evaluate which error type is the dominant factor. \\
\\
\textbf{I. Under-calling \& Scope Insufficiency} \\
* \textbf{Definition}: The agent fails to \textbf{initiate} the necessary tool calls to cover the full scope of the question. It makes incomplete attempts. \\
* \textbf{Scope}: The error is about \textbf{``What was NOT called''}. \\
* \textbf{Key Indicators}: \\
* \textbf{Direct Answering}: The agent answers the question directly (often hallucinating) WITHOUT calling any tools. \\
* \textbf{Partial Coverage}: The user asks for ``Director AND Actor'', but the agent ONLY calls a tool for ``Director'' (Missing scope). \\
* \textbf{Phantom Usage}: The agent claims to use a tool in thought, but no actual tool call is generated. \\
\\
\textbf{II. Tool Execution Failure} \\
* \textbf{Definition}: The agent attempts to use tools, but the \textbf{Tool Layer fails to provide usable data}, and the agent fails to recover. This covers both \textbf{Technical Failures} (Syntax) and \textbf{Data Failures} (Empty Results). \\
* \textbf{Scope}: The error is about \textbf{``The Tool Call yielded nothing useful''}. * \textbf{Key Indicators}: \\
* \textbf{Syntax/Schema Errors}: Persistent JSON errors, missing parameters, or wrong types that prevent execution. \\
* \textbf{Empty/Null Results}: The tool runs successfully but returns ``Not Found'', ``[]'', or ``None'', and the agent \textbf{cannot recover} (e.g., stops, loops, or gives up). \\
* \textbf{Unrecovered Mechanical Loop}: Repeatedly making the exact same failed call (Syntax or Empty) without changing strategy. \\
\\
\textbf{III. Reasoning Discontinuity} \\
* \textbf{Definition}: \textbf{Reasoning Process} breaks down. The logic connection between steps is flawed or incoherent. \\
* \textbf{Scope}: The error is about \textbf{Reasoning Discontinuity}. \\
* \textbf{Key Indicators}: \\
* \textbf{Context Loss / Interruption}: The agent starts a reasoning chain but abruptly stops or forgets previous constraints. \\
* \textbf{Logical Errors}: The agent forgets to the original question constraints, mixes up variables, or makes invalid deductions. \\
\\
--- \\
\textbf{Guidance on Identifying the ``Primary'' Error} \\
Use the following priority logic to decide the Primary one: \\
1. Check for Under-calling (Type I): \\
* Did the agent \textbf{fail to call} the necessary tools entirely? \\
* Did it miss a part of the question (e.g., checked date but missed location)? \\
* \textbf{If YES}, categorize as \textbf{I}. \\
2. Check for Tool/Retrieval Failure (Type II): \\
* Did the agent call the tool, but the tool \textbf{failed to work} (Syntax Error) OR \textbf{failed to return data} (Empty Result) and cannot recover? \\
* Did this failure cause the agent to get stuck, loop, or fail to produce an answer? \\
* \textbf{If all YES}, categorize as \textbf{II}. \\
3. Check for Reasoning Discontinuity (Type III): \\
* If agent dive in to reasoning but got lost or looped, mixed variables, or made illogical jumps? \\
* \textbf{If YES}, categorize as \textbf{III}. \\
\\
\textbf{Output Format} \\
\{ ``trajectory\_id'': ``ID or Summary'', ``category\_code'': ``Category ID (I, II, or III)'', ``category\_name'': ``Full Category Name'', ``reasoning'': ``Explain why this is the primary error type.'' \} Now, analyze the following trajectory: \\
\textbf{Question}: \{question\} \\
\textbf{Ground Truth}: \{ground\_truth\} \\
\textbf{Response Trajectory}: \{response\} \\
\hline
\end{tabular}
\caption{Prompt template for error type classification.}
\label{tab:error_judge_prompt}
\end{table*}

\end{document}